\documentclass[aps,prb,twocolumn,superscriptaddress,floatfix,eqsecnum]{revtex4}
\usepackage{amsmath}
\usepackage{amssymb}
\usepackage{wrapfig}
\usepackage{subfigure}
\usepackage{psfrag}
\usepackage{tabularx}

\usepackage[dvips]{graphicx}
\usepackage{feynmf}
\DeclareGraphicsExtensions{.eps}

\newcommand{\sign}{\,\mathrm{sign}\,}

\begin{document}
\title{Non-perturbative behavior of the quantum phase transition to a nematic
Fermi fluid}
\author{Michael J. Lawler}
\affiliation{Department of Physics, University of Illinois at Urbana-Champaign, 1110 W. Green Street, Urbana, Illinois 61801-3080, U.S.A.}
\author{Daniel G. Barci}
\affiliation{Departamento de F{\'\i}sica Te\'orica
Universidade do Estado do Rio de Janeiro.
Rua Sao Francisco Xavier 524,
20550-013,  Rio de Janeiro, RJ, Brazil.}
\author{Victoria Fern\'andez}
\affiliation{Department of Physics, University of Illinois at Urbana-Champaign, 1110 W. Green Street, Urbana, Illinois 61801-3080, U.S.A.}
\affiliation{Departamento de F\'{\i}sica, Universidad Nacional de
La Plata,
Argentina.}
\author{Eduardo Fradkin}
\affiliation{Department of Physics, University of Illinois 
at Urbana-Champaign, 1110 W. Green Street, Urbana, Illinois 61801-3080, U.S.A.}
\author{Luis Oxman}
\affiliation{Instituto de F\'{\i}sica, Universidade Federal Fluminense, Campus
da Praia Vermelha, Niter\'oi, 24210-340, RJ, Brazil. }

\date{\today}

\begin{abstract}
We discuss shape (Pomeranchuk) instabilities of  the Fermi 
surface of a two-dimensional Fermi system using bosonization. 
We consider in detail the quantum critical behavior of the transition 
of a two dimensional Fermi fluid to a nematic state which breaks 
spontaneously the rotational invariance of the Fermi liquid. 
We show that higher dimensional bosonization reproduces the 
quantum critical behavior expected from the Hertz-Millis analysis, 
and verify that this theory has dynamic critical exponent $z=3$. 
Going beyond this framework, we study the behavior of the fermion 
degrees of freedom directly, and show that at quantum criticality 
as well as in the the quantum nematic phase 
(except along a set of measure zero of symmetry-dictated directions)  
the quasi-particles of the normal Fermi liquid are generally wiped out. 
Instead, they exhibit short ranged spatial correlations that decay 
faster than any power-law, with the law 
$|x|^{-1} \exp(-\textrm{const.}\;|x|^{1/3})$ and we verify explicitely
the vanishing of the fermion residue utilizing this expression. In contrast, 
the fermion auto-correlation function has the behavior 
$|t|^{-1}  \exp(-{\rm const}.\;|t|^{-2/3})$.
In this regime we also find that, at low frequency, the  single-particle 
fermion density-of-states behaves as
$N^*(\omega)=N^*(0)+ B\; \omega^{2/3} \log\omega +\ldots$, where $N^*(0)$
is larger than the free Fermi value, $N(0)$, and $B$ is a constant.  These 
results confirm the non-Fermi liquid nature of both the quantum critical 
theory and of the nematic phase.
\end{abstract}

\maketitle
\section{Introduction}

The behavior of interacting Fermi systems near continuous 
quantum phase transitions is a central problem in the physics 
of strongly correlated systems. 
Although much work has been done on this subject, there are still many 
open and as yet unresolved questions. 
At present the standard theory of quantum phase 
transitions\cite{Hertz76,Millis93,Sachdev99} relies primarily on an 
analysis on the effects of fluctuations perturbatively about the results 
of Hartree-Fock theory. 
This analysis proceeds in almost complete analogy with the theory of 
classical critical phenomena about its upper critical dimension, 
and its straightforward extension to quantum phase 
transitions. In practice it consists of an effective theory for a 
suitable order parameter field while other degrees of freedom, including
fermions, are often integrated out at the outset.

In many cases of interest the systems are metallic and have 
gapless fermionic excitations. In the standard approach, their net effect is to 
introduce damping in the collective modes associated with the order parameter
field. In practice this results in the introduction of dissipative terms in the
effective action. While much of this is certainly correct, this approach
implicitly assumes that the fermions are largely unaffected by quantum
criticality.
Why this should be the case is far from obvious. 

The assumptions of the Landau theory of the Fermi 
liquid\cite{Pines66,AGD,baym91} are self-consistent and well justified 
within the Landau phase which has a sizable basin of stability, 
except in one dimensional 
systems\cite{shankar94,Polchinski94,Luther79,Haldane92,Haldane94,CastroNeto93,
CastroNeto94,CastroNeto95,Houghton93,Houghton00}. However, there is no reason 
for these assumptions to hold outside the Landau phase. However, there is also
 mounting evidence that these assumptions may also not hold in a 
 number of phases (and not just at quantum critical points), 
 including ferromagnetic
 metals\cite{Chubukov05} and nematic phases of Fermi 
 fluids\cite{Oganesyan01,Metzner03}.
The 
possibility that quantum
criticality may lead to non-Fermi liquid behavior has been 
a focus of research
in recent years, primarily (but not only) in connection with the physics of 
 the ``normal phase'' of high temperature 
 superconductors\cite{Chubukov94,Sachdev99,Varma96,Chakravarty01}, and with
 heavy-fermion systems\cite{heavy-fermions}.

The simplest example where the Landau assumptions on the behavior 
of the quasiparticles are violated is the quantum phase transition 
from a 
normal (Landau) Fermi liquid phase to a nematic Fermi fluid
\cite{Oganesyan01}. A nematic Fermi fluid is a uniform phase of a 
system of interacting fermions  in which 
the shape of the Fermi surface is distorted {\em spontaneously}, thus breaking 
rotational invariance.\cite{comment0} This state is an example of the fate of a Fermi liquid
beyond a Pomeranchuk 
instability\cite{Pomeranchuk58}. In this case, the Landau assumptions appear 
to be violated throughout this phase, 
and not just at the quantum critical point.\cite{Oganesyan01,Metzner03}

The clearest experimental evidence to date of a nematic Fermi fluid phase has been found
in very clean two-dimensional electron gases in magnetic fields in 
ultra-clean samples.\cite{Lilly99,Du99} 
The striking resistivity anisotropies
that are observed in these experiments can be explained by the
onset of nematic order at low temperatures.\cite{Fradkin99} 
It has also been proposed that phases of this type may play a central role 
on the physics of high temperature superconductors.\cite{Kivelson98,Kivelson03}
This  charge-ordered state of a strongly correlated system of fermions 
is the simplest example of an electronic liquid crystal phase\cite{Kivelson98}.

The problem of the fate of the fermions at quantum criticality, and in the 
``non-Fermi liquid'' phases mentioned above, so far has only been 
considered within perturbative corrections to Hartree-Fock/RPA theory. 
Oganesyan, Kivelson and Fradkin\cite{Oganesyan01} found that the 
quasiparticles are wiped out as well defined quantum states. This is due to 
the large fluctuations of (overdamped) quadrupolar collective modes. 
These authors found, within a Hartree-Fock and RPA theory, an
overdamped collective mode with $z=3$ at the critical point. 
They also found that the fermion self-energy acquires, 
at the quantum critical point, an imaginary part with a frequency 
dependence following the law $\omega^{2/3}$. 
A similar behavior has been found in the case of the Stoner transition
and in the antiferromagnetic phase\cite{Chubukov04,Chubukov05}.
Oganesyan and coworkers also found that this behavior holds 
inside the nematic phase, 
except along a set of measure zero of directions determined 
by the symmetry breaking.\cite{comment1}
However, it seems quite likely 
that such leading order behavior\cite{Oganesyan01,Metzner03} 
may actually signal the complete failure of the Landau theory of the Fermi 
liquid. It is clear that to better understand this problem 
a non-perturbative analysis of the behavior of the 
fermions at the quantum phase transitions (and beyond) is needed.
Chubukov\cite{Chubukov05} has given arguments which, in the context of the
ferromagnetic metallic transition, suggest that this behavior may persist beyond
the lowest order in perturbation theory. 

In this paper we will consider the nematic quantum phase transition in Fermi
fluids using the non-perturbative approach of higher dimensional 
 bosonization\cite{Haldane92,CastroNeto93,Houghton93}. We will not discuss
 the (important) lattice effects here.
 Bosonization is a powerful tool to study the non-perturbative
 behavior of one-dimensional
 gapless Fermi systems, the best understood fermionic quantum critical
 systems\cite{Emery79}. As it is well known, the kinematics of
 one-dimensional systems is so constrained that the bosonic collective modes
 completely exhaust the spectrum of these fermionic systems, allowing even for a
 full reconstruction of the fermionic operators entirely in terms of bosons. A
 striking result if one-dimensional system is that the electron acquires a
 non-trivial anomalous dimension and it is no longer the quasiparticle  of
 these systems, even for arbitrarily weak interactions. For these reasons
 one-dimensional gapless Fermi systems have been termed `Luttinger
 liquids'. The actual quasiparticles are non-trivial 
 solitons which are orthogonal to a bare electron.\cite{Haldane81}
 
In dimensions higher than one the physics (and the kinematics) 
is quite
different than in one dimension. Nevertheless bosonization methods still
yield 
the physics of the Landau theory of the Fermi
liquid correctly.\cite{Haldane92,CastroNeto93,Houghton93} Superficially this may seem
surprising since in dimensions higher than one there are no longer 
strong kinematic constraints, and consequently the bosonic collective modes 
cannot exhaust the spectrum of an interacting Fermi system. 
Instead, except
for narrow regimes in which the collective modes are stable quantum states, they
exhibit Landau damping, reflecting their decay into
particle-hole pairs. It is a key check of the validity of higher dimensional
bosonization that it gets the physics of Landau damping.\cite{CastroNeto94} 

One appealing feature of higher
dimensional bosonization is that it is actually a theory of the quantum 
fluctuations of the shape of the Fermi surface. It is thus a natural approach to
study quantum  phase transitions associated with Pomeranchuk instabilities, and
in particular the nematic state.\cite{Barci03} More specifically, we focus on 
the nematic case for spinless fermions and
compare with the work of Oganesyan and coworkers\cite{Oganesyan01} 
based on
RPA and Hartree-Fock. We find that the physics of the bosonic collective modes
is the same in bosonization and in RPA, and thus our results agree with those of
Ref.[\onlinecite{Oganesyan01}] in the Landau phase, in the nematic phase and at
the quantum critical point. Perhaps, this is not so surprising since at long
wavelengths RPA becomes asymptotically exact and this is the regime in which
bosonization is correct (for a more thorough discussion, see \onlinecite{Metzner98}). In particular we derive the effective action near the
quantum critical point and find that it does have a Hertz-Millis form with
dynamic critical exponent $z=3$, consistent with the findings Oganesyan and 
coworkers\cite{Oganesyan01}, 
and by Nilsson and Castro Neto\cite{Nilsson05}, but
in disagreement with the results of K. Yang\cite{Yang05}.

We further use bosonization methods to
obtain the fermion propagator. This result is well beyond the Hartree-Fock/RPA
theory and thus it allows us to study the fate of the fermions
non-perturbatively. We find striking violations of the Landau assumptions for
Fermi liquids. Thus, the equal-time behavior of the fermion propagator at the
quantum critical point (at zero
temperature) is found to fall off faster than any power, 
decaying instead with a law 
$\frac{1}{|x|}\; \exp(-{\rm const}. \; |x|^{1/3})$ as a function of
distance. The same behavior is found in the nematic phase except along 
symmetry-determined directions. We also verify explicitely from this expression the vanishing of the fermion residue as expected from this kind of behavior. In contrast
to the equal-time behavior, at quantum criticality the fermion auto-correlation 
function behaves as 
$\frac{1}{|t|} \; \exp(-{\rm const}.\; |t|^{-{2/3}})$, with a
similar albeit anisotropic law in the nematic phase as well. We also find that
the low energy behavior of the one-particle density of states $N^*(\omega)$ exhibits an
enhancement to a zero frequency value $N^*(0)$ which we find to be larger than $N(0)$, its non-interacting value. At finite but low frequency we further find that this one-particle density of states behaves as  $N^*(\omega)=N^*(0)+B\; \omega^{2/3} \ln \omega$ ($B$ is a constant), {\it i.e.\/} a cusp at $\omega=0$.

Thus, our bosonization results confirm that the nematic phase of a Fermi fluid is a {\em non-Fermi liquid}. However,
its behavior is more complex than the predictions of the Hartree-Fock/RPA theory. Recently Chubukov\cite{Chubukov05} has examined the behavior of the fermion self-energy in perturbation theory at the ferromagnetic quantum critical point and found that the frequency-dependence is not changed by higher order corrections. Our results for the auto-correlation function are  consistent with his results, as well as with Refs.[\onlinecite{Oganesyan01}] and [\onlinecite{Metzner03,Metzner05}]. However we also find that the equal-time
propagator (the ``one-particle density matrix'') has a very different behavior than what is predicted from these diagrammatic methods. 

The paper is organized as follows: In section
\ref{sec:pomeranchuk} we derive a theory of the nematic QCP via
higher dimensional bosonization. Here we present a theory of the quantum phase
transition to the nematic Fermi fluid, section 
\ref{p:saddle}, and show that it reproduces the analog of
Hertz-Millis theory for this
problem. In particular we give a detailed analysis of the spectral functions of
the collective modes,
section \ref{p:qmoments}, and derive the effective action valid 
in the vicinity of the quantum
phase transition, section \ref{p:otheory}. In section
\ref{sec:fermions} we use bosonization methods to calculate the fermion
propagator. Here we extract the full diagrammatic perturbation theory of the
fermion Green function from bosonization, and use it to calculate the fermion
self-energy. Here we 
check that the bosonization formulas reproduce correctly the non-Fermi liquid
behavior found within the Hartree-Fock/RPA theory\cite{Oganesyan01}.  
We then use the full bosonized 
 expression for
 the fermion propagator. Here we find large violations to Fermi liquid theory
 both at quantum criticality and in the nematic Fermi fluid phase. As an
 application we give a calculation of the fermion one-particle 
 density of states.
 Finally, in section \ref{sec:conc} we draw
our conclusions.  To help keep this paper self contained, in
Appendix \ref{ap:bosonintro} we give a short review the extension of
bosonization to $D$-dimensional Fermi systems. 
(For a more in depth review, see
Ref.[\onlinecite{Houghton00}]). 
In Appendix \ref{ap:part} we summarize details 
of the effective quadrupolar interactions, including fermion screening and 
Landau
damping effects. 
In Appendix \ref{ap:F0} we discuss the effects of the
(uncondensed) s-wave channel on the effective theory for the nematic. 
The details
of the calculation of the boson propagators are given in 
Appendix \ref{ap:equalt}.

\section{The nematic quantum phase transition and the order parameter}
\label{sec:pomeranchuk} 

In this section, we consider the boson
theory, obtained via an extension of bosonization to greater than
one dimension, near a nematic (Pomeranchuk) instability of a translationally
invariant fermion system. In the notation of Appendix
\ref{ap:bosonintro}, we take the following action for the bosons
\begin{multline}
\label{eq:baction0}
  S_0 = \frac{N(0)}{2}\sum_S\!\int d^2xdt\bigg[-\partial_t\varphi_S{\bf v}_S
  \cdot\nabla\varphi_S-\big({\bf v}_S\cdot\nabla\varphi_S\big)^2\bigg]
\end{multline}
and forward scattering interactions
\begin{widetext}
\begin{equation}
\label{eq:bactionInt}
  S_{\rm int} =
   \frac{N(0)}{2}
   \sum_{S,T}\int d^2xd^2x'dt\;
  F_{S-T}({\bf x-x'})
  {\bf v}_S\cdot\nabla\varphi_S({\bf x}){\bf v}_T\cdot\nabla\varphi_T({\bf x'})
\end{equation}
\end{widetext}
Here, $S$ labels the patch defined by coarse graining the Fermi
surface, and the density of quasiparticles in a patch may be
obtained from the boson field $\varphi_S({\bf x},t)$ via the
relation $\delta n_S({\bf x},t) = N(0){\bf
v}_S\cdot\nabla\varphi_S({\bf x},t)$. $F_{S-T}({\bf x}-{\bf x}')$
is therefore the interaction between particle-hole pairs in
patches $S$ and $T$.

We begin by analyzing our bosonized theory for a constant field configuration
and reproduce Pomeranchuk's result. Consider
configurations such that $\delta n_S$ is constant in space and
time over some particular range of time T. The resulting action
is:
\begin{multline}
  S_{\text{shape}} = -\frac{VT}{N(0)}\left(1+F_0\right)\left(m_0^+\right)^2 \\
    -\frac{VT}{2N(0)}\sum_{\ell>0}^{N/2}\left(1+F_\ell\right)
    \left(\left(m_\ell^+\right)^2 + \left(m_\ell^-\right)^2\right)
\end{multline}
where we have expanded $\delta n_S$ as:
\begin{equation}
\label{eq:defml}
  \delta n_S = \sqrt{\frac{2}{N}}\sum_{\ell=0}^{N/2} \left[m_{\ell}^+
  \cos(\ell\theta_S)
     + m_{\ell}^-\sin(\ell\theta_S)\right]
\end{equation}
and introduced the Fermi liquid parameters via
\begin{equation}
\label{eq:FST}
  F_{S-T} = \frac{1}{N} F_0 +
    \frac{2}{N}\sum_{\ell>0}F_\ell\cos\ell\left(\theta_S-\theta_T\right)
\end{equation}
Hence, for arbitrary $m_\ell^\pm$, we find that any $F_\ell<-1$
will destabilize the Fermi liquid. The point $F_\ell=-1$ we shall
call the Pomeranchuk (nematic for $\ell=2$) quantum critical point (QCP). Though
Fermi liquid theory breaks down at this point, Luttinger's
theorem is still obeyed.

It should also be noted that in the above analysis we could have
included interactions involving large angle scattering (which may
lead to charge density / spin density wave instabilities),
corrections to the linearized dispersion, three- or four-body
interactions and BCS processes in the bosonized theory. However,
except for BCS processes, these effects are irrelevant in the
Fermi liquid phase though some become important near the the
nematic critical point to be discussed below.

For simplicity, in the rest of this section, we shall specialize
to the $\ell=2$ instability in two-spatial dimensions, the 2D
quantum nematic liquid crystal, though the results generalize
easily.

\subsection{Saddle point expansion near the nematic QCP}
\label{p:saddle}

To study the nematic QCP, originally considered by Oganesyan, Kivelson and Fradkin\cite{Oganesyan01}, we set all $F_{\ell}$ to zero for $\ell\neq2$. This is
reasonable since these other modes are not critical and their
effect is only to introduce finite renormalizations of the
parameters in the effective theory of the critical (quadrupolar)
modes (see below).

On the broken symmetry side (i. e. $F_2 < -1$), the quadratic
action is no longer stable. So, in order to make the theory
consistent, we need (at least) a quartic term in the bosonized
action. Here, as a specific example, we consider the quartic
interaction that arises from corrections to the linearized
dispersion in the bosonized form given in Ref.[\onlinecite{Barci03}],
\begin{equation}
 S_{4} = \frac{\gamma N(0)}{4!} \sum_{S} \int d^2 r dt
    \big(\mathbf{v}_S\cdot\nabla\varphi_S\big)^4 
    \label{quarticaction}
\end{equation}
This term can be found by a direct extension of our $\epsilon_n$
expansion of \eqref{eq:pointsplit} to third order, noticing that
it is an exponential series in $i\delta n_S\epsilon_n/(N(0)\hbar
v_F)$. An estimate of $\gamma$ may therefore naturally be obtained
from the equations of motion of either the boson or fermion
pictures. However, this is not the only quartic contribution to
the action since an 8-fermion interaction term (that is quartic in
densities) also contributes at this level with a more general form
similar to the Fermi liquid interactions. Nevertheless, this will
give us a flavor of the broken symmetry phase.

Since this effective theory is no longer quadratic in the
bosonic fields, we will examine its behavior within a semiclassical
approximation, which means that we will first find the extremal
configuration and then expand about it. Thus, we write
\begin{equation}
 \delta n_S = N(0){\bf v}_S \cdot \nabla \varphi = \delta n^{cl}_S + \xi_S
\label{changefield}
\end{equation}
where $\delta n_S^{cl}$ is the solution of the classical equations of
motion at a uniform, mean field level:
\begin{multline}
\delta n^{cl}_{S} +
  \frac{2}{N}F_2(0)\sum_{T} \cos\big(2(\theta_S-\theta_T)\big) 
  \delta n^{cl}_{T} \\+
  \frac{\gamma}{3!N(0)^2} \big(\delta n^{cl}_{S}\big)^3 = 0
\end{multline}
where as stated earlier, we set $F_\ell(0)=0$ for
$\ell\neq2$. Because the non-linear term is cubic, we seek
solutions of the type
\begin{equation}
\delta n^{cl}_S = 
\frac{1}{\sqrt{N}}\sum_{\ell=-N/2}^{+N/2}m_\ell e^{i\ell\theta_S}
\end{equation}
where $m_\ell\neq0$ for $\ell=\{\pm2, \pm6, \pm10, \ldots\}$. In
terms of the $m_\ell$'s, the equation of motion becomes
\begin{multline}
m_{\ell}+F_2 (m_2 \delta_{\ell,2} + m_{-2} \delta_{\ell,-2}) +\\
\frac{\gamma}{3!N(0)^2N} \sum_{\ell_1,\ell_2}
m_{\ell_1}m_{\ell_2}m_{\ell-\ell_1-\ell_2} = 0
\label{recurrence}
\end{multline}
We first observe that for $F_2\geq -1$, the only possible
solution is $m_\ell=0$ for all $\ell$, as expected for the
isotropic case. If $F_2 <-1$, there exits a whole set of non
trivial solutions involving, in general, all harmonics
$\ell=\{\pm2,\pm6,\pm10,\ldots\}$ obeying particle hole symmetry.
Nevertheless, when we are near the phase transition, {\it i. e.\/}
$F_2\lesssim -1$, one can find the set of solutions analytically
with:
\begin{equation}
   m_2m_{-2} = \frac{1}{2}\left(\left(m_2^+\right)^2 +\left(m_2^-\right)^2\right)
     = \frac{2}{\gamma}N(0)^2N\mid 1+F_2(0)\mid
\end{equation}
using the notation of \eqref{eq:defml} and for the higher harmonics
\begin{equation}
  \mid m_{4k+2}\mid^2\propto
    N(0)^2N\left(\frac{2}{\gamma}\mid 1+F_2(0)\mid\right)^{2k+1}
\end{equation}
so that in the limit $F_2\to -1$ we can neglect the higher harmonics.

From this calculation, we conclude that near the ~ ~ $F_2=-1$ nematic
QCP, the Fermi surface takes the simple shape:
\begin{equation}\label{eq:fsshape}
  \delta n_S^{cl} = N(0)\sqrt{\frac{2|1+F_2(0)|}{\gamma}}\cos(2\theta_s-\alpha)
\end{equation}
where $\alpha$ picks out the major axis of the ellipse that is
spontaneously chosen. Without loss of generality, from now on we
will set $\alpha=0$. We also conclude that further away from the
critical point, the anharmonic quartic terms generically introduce
higher harmonics to this shape. From this point of view for
example, the Fermi surface in the nematic phase may become
increasingly flatter away from the critical point leading to an
additional instability towards a smectic phase breaking
translational order in one direction. However, this phase
transition cannot be seen within our Forward scattering only model
and all that happens deeper into the phase here is that the shape
becomes more anharmonic. It should also be noted that the other
harmonics, $\ell=\{0,\pm4,\pm8,\ldots\}$, appear when particle
hole symmetry is broken, {\it e. g.\/} when a cubic term in the action is
introduced ($S_3$) corresponding to the addition of quadratic
terms to the fermion dispersion.

\subsection{Theory of the Quadrupole Moment Density}
\label{p:qmoments}

The next step beyond mean field theory is to formulate an order
parameter theory. To that end, in this section we are interested
in focusing on describing the behavior of the quadrupole moment
density, $m_2^{\pm}({\bf q},\omega)$. Again, for our present
purposes we shall keep only $F_2(q)$. However, in Appendix \ref{ap:F0} we
show that by keeping also $F_0(q)$, the additional effects of the non-critical 
modes do not change our results in any essential way.

Now, our goal here is to obtain an action entirely in terms of
$m^{\pm}_{2}$. This can be easily accomplished by following
Ref.[\onlinecite{Houghton00}] and using a Hubbard-Stratonovich 
transformation
to aid the diagonalization. In the end, however, both the
auxiliary fields and $m^{\pm}_{\ell}$ for $\ell\neq 2$ shall be
integrated out.

In momentum space, our free action can be written as
\begin{equation}
\label{eq:freebosonS}
  S_0 = \frac{1}{2}\sum_S\int\frac{d^2qd\omega}{(2\pi)^3}
    \big(\chi^0_{S}\big)^{-1}(q,\omega)
    \delta n_S({\bf q},\omega)\delta n_S(-{\bf q},-\omega)
\end{equation}
where
\begin{equation}
  \chi^0_{S}({\bf q},\omega) =
    N(0)\frac{{\bf v_S}\cdot{\bf q}}{\omega -{\bf v_S}\cdot{\bf q}}=
    N(0)\frac{\cos\theta_S}{s - \cos\theta_S}
\end{equation}
is the density-density response function in the ``small q'' limit
of patch $S$ with $s=\omega/qv_F$. (Note: when needed, we
regularize the denominator by letting $s\to s+i\epsilon\sign(s)$
according to the usual time ordering prescription.) 

The interactions, described by the quadratic action $S_{\rm int}$, {\it c.f.\/} 
Eq.\eqref{eq:bactionInt}, become {\em diagonal} in the angular momentum basis,
{\it i.e.\/} in terms of the multipole densities $m_\ell^\pm(q,\omega)$. In
particular, the contribution to $S_{\rm int}$ from the $\ell=2$ (quadrupolar) 
densities, is
\begin{equation}
  S_{\rm int} = \frac{1}{2}\int\frac{d^2qd\omega}{(2\pi)^3} f_2(q) 
  \bigg[\mid m_2^+(q,\omega)\mid^2 +
    \mid m_2^-(q,\omega) \mid^2\bigg]
\end{equation}
with $f_2(q)$ defined as usual in Fermi liquid theory through
$F_2(q) = N(0)f_2(q)$. The contributions from the other angular momentum
channels  have a similar form (in terms of the respective Landau parameters.)

To aid the diagonalization, we split up the free part of the action, 
$S_0$, using the
Hubbard-Stratonovich transformation:
\begin{multline}
  S_0 = -\frac{1}{2}\sum_S\int\frac{d^2qd\omega}{(2\pi)^3}\bigg[
    \chi^0_{S}(q,\omega)\sigma_S(q,\omega)\sigma_S(-q,-\omega) +\\
    \big(\sigma_S(q,\omega)\delta n_S(-q,-\omega) +
    \sigma_S(-q,-\omega)\delta n_S(q,\omega)
    \big)\bigg]
\end{multline}
and then switch over to the angular momentum basis:
\begin{multline}
  S_0 = -\frac{1}{2}\sum_{\eta=\pm}\sum_{\ell,\ell'}
        \int\frac{d^2qd\omega}{(2\pi)^3}\\
    \bigg[ 
    \big(\chi^0_{\ell-\ell'}+\eta\chi^0_{\ell+\ell'}\big)
    \sigma_{\ell}^{\eta}\sigma_{\ell'}^{\eta} +
    2(\delta_{\ell-\ell',0}+
    \eta\delta_{\ell+\ell',0})\big(\sigma_{\ell}^{\eta}m_{\ell}^{\eta}
    \big)\bigg]\label{eq:fullsigma}
\end{multline}
In the large-$N$ limit, we have
\begin{equation}
  \chi^0_\ell = N(0)\int_0^{2\pi} \frac{d\theta}{2\pi}
    \frac{\cos\theta}{s-\cos\theta}e^{i\ell\theta}
\end{equation}
Here we have Fourier transformed with respect to 
$\hat{\bf q}=(\cos\phi,\sin\phi)$, that is, $\theta=\theta_S-\phi$.

Now, we will integrate out all the $m_{\ell}^{\pm}$ densities, except for
$\ell=2$. This can be easily done since $S_0$ is a linear function of these
fields, while they are absent in $S_{\rm int}$. This is so for this model
with only a quadrupolar interaction , {\it i.e.\/} we have set their
corresponding Fermi liquid parameters to zero. (In the vicinity of the nematic
transition it is straightforward to include
the effects of the $\ell\neq 0$ channels. Their net effect is to give rise to
simple renormalizations of the effective theory we are about to derive. A detailed analysis is given 
Appendix \ref{ap:F0}).
The result is a delta function for the $\sigma_{\ell}$ fields, 
allowing us to also
integrate them out with the net result to simply set
$\sigma_{\ell}^{\pm}=0$ for all $\ell$ except $\ell=2$. This gives
us the following simple expression for the effective free action:
\begin{multline}
  S_0 = -\frac{1}{2}\sum_{\eta=\pm}\int\frac{d^2qd\omega}{(2\pi)^3}
    \bigg[ 
    \big(\chi^0_{0}+\eta\chi^0_{4}\big)
    \big|\sigma_{2}^{\eta}\big|^2 +
    2\big(\sigma_{2}^{\eta}m_{2}^{\eta}
    \big)\bigg]
    \\
\end{multline}
and we now have an action entirely in terms of the quadrupole moment density.

The final step is to integrate out the $\sigma^{\pm}_{2}$ fields. This is
easily accomplished and we obtain the Gaussian level of the order
parameter theory, including the effects of the interactions $S_{\rm int}$. 
The action of the effective theory is:
\begin{equation}
\label{eq:S2}
  S_{2}[m_2^+,m_2^-]= \frac{1}{2}\int\frac{d^2qd\omega}{(2\pi)^3}\left[
  \frac{1}{\chi_2^{+}}\big|m_2^{+}\big|^2 +
  \frac{1}{\chi_2^{-}}\big|m_2^{-}\big|^2  \right]
\end{equation}
where $\chi_2^\pm(s,q)$ is the dynamical correlation function (susceptibility)
of the quadrupolar densities:
\begin{equation}
  \chi_2^{\pm}(s,q)=\frac{\chi^0_0(s) \pm \chi^0_{4}(s)}{
    1-f_{2}(q)\big(\chi^0_0(s) \pm \chi^0_{4}(s)\big)}
\end{equation}
and
\begin{equation}
  \chi^0_{2\ell} = N(0)\bigg[-\delta_{\ell,0} +
    K_0(s)\bigg(\frac{1-K_0(s)}{1+K_0(s)}\bigg)^\ell\bigg]
    \label{eq:chi0}
\end{equation}
with 
\begin{equation}
K_0(s)=\frac{s}{\sqrt{s-1}\; \sqrt{s+1}}
\label{eq:K0}
\end{equation} 
Naturally, this is just RPA quadrupolar susceptibility of 
Ref. [\onlinecite{Oganesyan01}]. 

We should stress that the effective action $S_2$ of Eq.\eqref{eq:S2} {\em
does not} include the effects of the non-linear interactions represented 
in $S_4$, {\it c.f.\/} Eq.\eqref{quarticaction}, which are crucial to 
stabilize the nematic
state past the nematic QCP (for $F_2(0)<-1$). 
As we discussed above, these
non-linear terms do mix the different angular momentum channels. However,
provided there is no condensation for $\ell \neq 2$ it is still possible to
integrate out these degrees of freedom, at least perturbatively. Thus, 
sufficiently close to the nematic QCP, the effects of the higher 
angular momentum channels will remain perturbatively small.

Furthermore, should we have decided to integrate out the density
fields, $m_2^\pm$ in favor of the auxiliary fields
$\sigma_2^\pm$, we would have found the propagators of the
$\sigma_\ell^\pm$ fields to be the RPA effective interaction:
\begin{align}
  V^{\pm}_2(s,q) &= \frac{f_2(q)}{1 -
     f_2(q)\big(\chi^0_0(s) \pm \chi^0_{4}(s)\big)}
\end{align}
The action of the $\sigma_\ell^\pm$ fields is precisely that
found by Oganesyan and coworkers\cite{Oganesyan01} also obtained through a
Hubbard-Stratonovich transformation but performed directly in the
fermion theory. We shall find that understanding this interaction
is the key to understanding the physics of the nematic QCP.

\subsection{Order parameter theory of the nematic QCP}
\label{p:otheory}

The theory with the action given by Eqs.\eqref{eq:baction0} and \eqref{eq:bactionInt} seemingly describes the quantum mechanics of a fluctuating surface. This suggests that the effective degrees of freedom ought to be long-lived bosonic modes of the fluctuations of the shape of the Fermi surface. In other terms, this bosonized theory would seem to be entirely described by stable collective modes. However, these bosonic excitations are not generally stable due to {\em Landau damping} effects, represented by the branch cut singularities  in \eqref{eq:chi0}\cite{Oganesyan01,Chubukov04b}. 

We will examine this problem more closely. It is useful to introduce 
 the {\em quadrupole density spectral functions}:
\begin{equation}
\label{eq:spect}
\begin{split}
  S_2^\pm(q,s) &\equiv -2\sign(s){\mathcal Im}\chi_2^\pm(q,s)\\
   &=-\frac{2\sign(s){\mathcal Im}V_2^\pm(q,s)}{f_2(q)^2}
   \equiv \frac{B_2^\pm(q,s)}{f_2(q)^2}
\end{split}
\end{equation}
where $B_2^\pm(q,s)$ denotes the spectral function of $V_2^\pm(q,s)$, 
the correlation function of the $\sigma_2^\pm$ fields.

The analysis is greatly simplified upon recognizing that
$V_2^\pm(q,s)$ is a polynomial function of  $K_0(s)$. For $\ell=2$ the
denominator of $V_2^+(q,s)$ is a cubic function of $K_0$, whereas for
$V_2^-(q,s)$ it is quadratic in $K_0$.
In Appendix \ref{ap:part} we show that $V_2^\pm(q,s)$ have the partial fraction
expansions:
\begin{equation}
\label{eq:Vexpand}
  V_2^\pm(q,s) = \frac{1}{N(0)}\sum_{\beta}
  \frac{{\mathcal Z}^\pm_\beta(q) }{\delta^\pm_\beta(q)-K_0(s)}
\end{equation}
where $\beta=a,b,c$ for $V_2^+$, and $\beta=a,b$ for $V_2^-$.
We can therefore view $V_2^\pm(q,s)$ as a sum of terms each of the form of
the RPA s-wave channel effective interaction,
renormalized by a residue ${\mathcal Z}_\beta^\pm$ and with an
effective interaction $f_\beta^\pm(q)=(1-\delta^\pm_\beta(q))^{-1}$. 
Details of these
expansions are given in Appendix \ref{ap:part}. 

\begin{figure}[!t]
\psfrag{S}{$\!\!\!\!\!\!\!\!\!\!\!\!\!\!\!\!\!\!\!
\widetilde S_2^+(q,s)$}
\psfrag{s}{$s$}
\psfrag{30}{}
\psfrag{25}{}
\psfrag{20}{}
\psfrag{15}{}
\psfrag{10}{\tiny $\!\!\!\!\!\! 10$}
\psfrag{5}{}
\psfrag{0.2}{}
\psfrag{0.4}{}
\psfrag{0.6}{}
\psfrag{0.8}{}
\psfrag{1}{\tiny $\!\!\!\!\!\! 1$}
\includegraphics[width=0.45\textwidth]{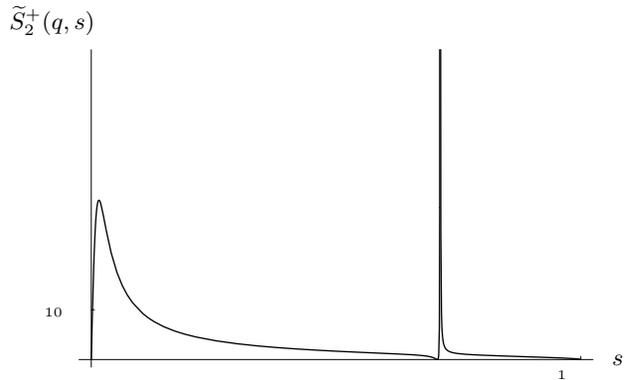}
\caption{The (normalized) quadrupole density spectral function 
${\widetilde S_2^+}(q,s)=S_2^+(q,s)/N(0)$, Eq. \eqref{eq:spect} and 
Eq.\eqref{eq:V2+} for $q \neq 0$, very close to the nematic quantum 
phase transition from 
the Fermi liquid phase, $F_2(0) \to -1^+$. 
As the QCP is approached,
 there is a large increase of spectral weight in the 
 the $z=3$ overdamped quadrupolar mode
 at very low frequency. Notice the delta function contribution of 
 the propagating mode with $z=1$ discussed in the text. The propagating mode
 with $z=2$ gives a similar delta function contribution 
 (with smaller spectral weight) to
 $S_2^-(q,s)$.}
\label{fig:V}
\end{figure}
Near the nematic
QCP, where $|F_2(0) +1|\ll 1$, upon defining the quantity $\delta_2(q)$
\begin{equation}
\label{eq:delta2}
\delta_2(q)=\frac{1+F_2(q)}{F_2(q)}= 1+\frac{1}{F_2(0)}-\kappa
q^2,
\end{equation}
 the effective interactions $V_2^\pm(q,s)$ 
 become simple and we obtain:
\begin{eqnarray}
&&\!\!\!\!\!\!\!\!\!\!\!\!\!\!\!\!\!\!\!\!\!\!\!\!
 V_2^+(q,s) \approx \frac{1}{2N(0)}\bigg[\frac{1}{\frac{\delta_2(q)}{2}-K_0(s)}
    - \frac{1/4}{s^2-1/2}\bigg]
    \label{eq:V2+}\\
&&\!\!\!\!\!\!\!\!\!\!\!\!\!\!\!\!\!\!\!\!\!\!\!\!
V_2^-(q,s) \approx \frac{1}{4N(0)}\frac{1}{s^2+\frac{\delta_2(q)}{4}}
\label{eq:V2-}
\end{eqnarray}
Thus, near the critical point we find a propagating mode
with dispersion $\omega_q=\left(1/\sqrt{2}\right)qv_F$ (a $z=1$
mode), represented by the pole in the second term for $V_2^+$ 
(see Eq.\eqref{eq:V2+}), and another one with dispersion
$\omega_q=\left(\sqrt{\delta_2(q)}/2\right)qv_F$ (a $z=2$ mode), given by the
pole in $V_2^-$ (see Eq.\eqref{eq:V2-}). Furthermore, we also find an overdamped mode, 
given by the pole in first term of $V_2^+$ (see
Eq.\eqref{eq:V2+}),
with a dispersion relation of
$s=-i\delta_2(q)/2$, and dynamic critical exponent $z=3$. 
For $q$ sufficiently small, and asymptotically close 
to the nematic quantum critical 
point, we find that this overdamped mode
dominates the spectral function over the other two (propagating) modes.

In Fig. \ref{fig:V} we present a plot the spectral
function ~~ $B_2^+(q,s)=-2\sign(s){\mathcal Im} V_2^+(q,s)$.
It shows that, as the QCP is approached from the 
Fermi liquid side,
there is a large transfer of spectral weight in the  
quadrupolar spectral
function to the low frequency end of the spectrum, associated with the
 emergence of the overdamped $z=3$ mode. 
 This mode thus controls the quantum
 critical behavior.
 
Conversely, normal Fermi-liquid
behavior is obtained if $\delta_2(0)$ is finite so that $z=1$. 
On the other hand, if
the overdamped mode were to be absent, the dynamic critical behavior would be
controlled by propagating mode with $z=2$ discussed above.
The trend is thus opposite to what one might naively expect: the
higher the $z$, the higher the effective dimension $D+z$, the
stronger the  divergence of the overdamped mode. Recently, Yang\cite{Yang05}
proposed that the transition to the quantum nematic state should have dynamic
critical exponent $z=2$. The analysis we just presented shows that this is not
the case.

As a result of the above analysis, we obtain the following 
effective action
for the quadrupole density near the nematic QCP:
\begin{multline}
  S_{\rm QCP} = \frac{1}{2N(0)}\int \frac{d^2qd\omega}{(2\pi)^3}\bigg[
    \big(2i|s| - 1-F_2(q)\big)|m^+_2|^2\\ +
    \big(4s^2 - 1-F_2(q)\big)|m^-_2|^2]\bigg] \\-
    \frac{\gamma}{8N(0)^3N}\int d^2xdt
    \big({m^+_2}^2+{m^-_2}^2\big)^2
\end{multline}
The order parameter field is
\begin{equation}
\label{eq:m2}
\begin{split}
  m^+_2({\bf q},\omega) &=
  \sqrt{\frac{2}{N}}\sum_S\delta n_S({\bf q},\omega)\cos2(\theta_S-\phi)\\
  m^-_2({\bf q},\omega) &=
  \sqrt{\frac{2}{N}}\sum_S\delta n_S({\bf q},\omega)\sin2(\theta_S-\phi)
\end{split}
\end{equation}
\noindent where, again, $\phi$ is the direction of ${\bf
q}=q(\cos\phi,\sin\phi)$. This action is therefore very similar to
Hertz's action for the ferromagnetic quantum phase transition in itinerant
fermionic systems\cite{Hertz76},
 but here
within the context of the formation of nematic order (and similar
actions may also be obtained for higher $\ell$). Recently,
Nilsson and Castro Neto \cite{Nilsson05} derived this action using Fermi liquid
theory methods.

Now, let us look on the broken symmetry side, in the nematic
phase. As in an in any theory with an $O(2)$ symmetry, 
here we will find that the order
parameter will spontaneously pick a direction. As a result, it is
no longer useful to Fourier transform with respect to $\phi$ in Eq.
\eqref{eq:m2}. Rotating back and after the saddle point expansion
about the classical configuration \eqref{eq:fsshape}, we find the
quadratic action on the nematic side:

\begin{multline}
  S_{\rm QCP}^{\rm nematic} = \frac{1}{2N(0)}\int \frac{d^2qd\omega}{(2\pi)^3}
    {\bf m}_2({\bf q},\omega)
    \cdot{\bf \chi}_2^{-1}\cdot{\bf m}_2(-{\bf q},-\omega)
\end{multline}
where $\textbf{m}_2(\textbf{q},\omega)=(m_2^+(\textbf{q},\omega),m_2^-(\textbf{q},\omega))$ and
\begin{multline}
  {\bf \chi}_2^{-1} = (i|s|-\kappa q^2-|1+F_2(0)|)
  \begin{pmatrix}1&0\\0&1\end{pmatrix}\\
  + i|s|
\begin{pmatrix}
  \cos4\phi & \sin4\phi\\
  \sin4\phi & -\cos4\phi
\end{pmatrix}
 -|1+F_2(0)|
\begin{pmatrix}
  1 & 0\\
  0 & -1
\end{pmatrix}
\end{multline}

\noindent in the reference frame in which the nematic 
order parameter is diagonal,
{\it i.e.\/} its principal axes, whose orientation is determined 
spontaneously.  The difference between this action and
the previous one discussed above for the symmetric phase, 
is the emergence of the
last term which originates from the non-linear (quartic) 
term in the effective 
action.  This term ruins our ability to rotate $\phi$ out
of the action. 

Noting that the off-diagonal terms are higher order
in $s$, we may write this in the simplified form:

\begin{widetext}
\begin{equation}
  S_{\rm QCP}^{\rm nematic} = \frac{1}{2N(0)}\int \frac{d^2q}{(2\pi)^2}
  \int \frac{d\omega}{2\pi}
  \bigg[
  \big(2i|s|\cos^2(2\phi) -2|1+F_2(0)|-\kappa q^2\big)|m^+_2|^2
  + \big(2i|s|\sin^2(2\phi)-\kappa q^2\big)|m^-_2|^2\bigg]
  \label{eq:Sqcp-nematic}
\end{equation}
\end{widetext}

We see that now $m_2^+$ is the amplitude mode, which has $z=1$, while
$m_2^-$ is the nematic Goldstone mode \emph{which continues to have 
dynamic critical
exponent $z=3$ even in the nematic phase}.\cite{Oganesyan01}

The last point is to discuss
how this affects the original boson theory. If we bring back all
the integrated out angular momentum channels, we find that the
free action on the broken symmetry side has:
\begin{equation}
  \widetilde{\chi}^0_{S}({\bf q},\omega) =
    N(0)\frac{{\bf v_S}\cdot{\bf q}}{\omega -
    {\bf \widetilde{v}_S}\cdot{\bf q}}
\end{equation}
with a weakly renormalized Fermi velocity
\begin{equation}
  {\bf \widetilde{v}}_S = \big(1+4|1+F_2(0)|\cos^2(2\theta_S)\big){\bf v}_S
\end{equation}
Hence, spontaneous symmetry breaking essentially produces a
Goldstone mode that continues the critical, $z=3$ behavior into
the nematic phase while leaving the rest of the theory virtually
untouched until deep into the broken symmetry phase.

\section{fermions in the critical regime}
\label{sec:fermions}

In the past sections we discussed the behavior of the collective modes near the
nematic quantum phase transition. 
Much of what we discussed in the
previous section on the behavior of the collective modes is indeed 
in complete agreement
with the RPA treatment of this theory\cite{Oganesyan01}. This should not be a
surprise since RPA is asymptotically exact at low energies and at low
frequencies. This is also the reason while bosonization works in the same regime.

We will now turn our attention to the behavior of the fermionic degrees of freedom
near the nematic QCP and and in the nematic phase. This is very different problem. 
In
Ref.[\onlinecite{Oganesyan01}] the behavior of the fermion Green function was
studied perturbatively and a startling non-Fermi liquid behavior 
was found already at the lowest (``Fock'') order. However, this very finding raises
questions on the applicability of perturbation theory for the fermion propagator. In
this section we will use bosonization methods to address this problem.

Here we will use bosonization to compute the fermion propagator. Within this approach one has a theory for the bosonized degrees of freedom and a set of operator identities relating observables of the fermionic theory to those of the bosonic theory.  For a summary see Appendix \ref{ap:bosonintro}. There are two important issues to keep in mind. One is that the bosonized theory is exact for a fermionic theory with a linearized dispersion and forward scattering interactions ({\it i.e.\/} those described by Landau parameters). The other is that one has, within this theory, an operator to represent the fermion. Corrections to the linear dispersion as well as other (non-forward scattering) interactions are represented by non-linear terms in the action of the bosonized theory. The expressions that we will derive below apply strictly speaking to the fixed point theory, in which these perturbations are not included. It turns out that corrections due to the non-linearities of the fermion dispersion and other such terms are irrelevant (both in the Landau phase and at the quantum critical point). As such they will affect the results at high energies and at momenta but their effects become negligible in the low energy limit. Please note that one such operator, discussed in the previous section, stabilizes the nematic phase, {\it i.e.\/} it is a prototypical dangerous irrelevant operator.

The boson Green function may be found from our action in a similar
way as we found the above density-density correlation functions,
that is, using a Hubbard-Stratonovich approach. The result is:
\begin{widetext}
\begin{equation}
\label{eq:GB}
  G_{B(S,T)}({\bf x},t) = G_{B(S,T)}^0({\bf x},t) +
    i\int\frac{d^2kd\varepsilon}{(2\pi)^3}G^0_{F(S)}(k,\varepsilon) 
    V_{S,T}(k,\varepsilon)
    G^0_{F(T)}(k,\varepsilon)
    \left(e^{i({\bf k}\cdot{\bf x} - \varepsilon t)}-1\right)
\end{equation}
\end{widetext}
where $G_{B(S,T)}^0$ is infrared divergent unless ${\bf
x}\parallel {\bf v}_S$ and $S=T$. On the same patch, $G_{B(S,S)}^0({\bf x},t)$ is
given by the standard expression
\begin{eqnarray}
G_{B(S,S)}^0({\bf x},t)&&=\langle \varphi_S({\bf x},t) \varphi_S({\bf 0},0)\rangle-
\langle \varphi_S({\bf 0},0)^2 \rangle
\nonumber \\
&&=-\ln \left(\frac{\hat n_S \cdot {\bf x}+ i v_F t + i a \sign t}{ia}\right)
\nonumber \\
&&
\end{eqnarray}
where $a$ is a short-distance cutoff. 
In Eq.\eqref{eq:GB} we have denoted by $V_{S,T}(q,\omega)$ the effective interaction
\begin{multline}
  V_{S,T}(q,\omega) = \\
  \displaystyle{\frac{1}{N(0)}}\sum_{\ell,\ell'}
    e^{i(\ell (\theta_S-\phi) +\ell'(\theta_T-\phi))}
    \langle\sigma_\ell(q,\omega)\sigma_{\ell'}(-q,-\omega)\rangle
    \\
\end{multline}
which for the quadrupolar case, becomes simply
\begin{multline}
\label{eq:VST}
 V_{S,T}(q,\omega) = V^+_2(q,\omega)\cos2(\theta_S-\phi)\cos2(\theta_T-\phi) 
 \\+
 V^-_2(q,\omega)\sin2(\theta_S-\phi)\sin2(\theta_T-\phi)\\
\end{multline}
where the relative angle, $\theta_S-\phi$ appears here (see Eq.
\eqref{eq:fullsigma}). We also have used in \eqref{eq:GB}
the free fermion Green function:
\begin{equation}
  G^0_{F(S)}(q,\omega) = \frac{1}{\omega - {\bf v}_S\cdot{\bf q}+
                       i\epsilon\sign(\omega)}
\end{equation}
We will now use the bosonized form of the fermion operator 
(see Eq. \eqref{eq:psi} of Appendix \ref{ap:bosonintro})
to give us the fermion Green
function of the form:
\begin{align}
  G_F({\bf x},t) &= \sum_{S,T}\;  \left(\frac{-i}{N}\right)\;
    \langle T \; \psi_S({\bf x},t)\psi_T^\dagger({\bf 0},0) \rangle
    e^{i{\bf k}_S\cdot{\bf x}}\\
  &\equiv \frac{1}{N} \sum_S  G_{F(S)}({\bf x},t)e^{i{\bf k}_S\cdot{\bf x}}
\end{align}
for which we find the explicit expression
\begin{widetext}
\begin{equation}
  G_{F(S)}({\bf x},t) = G_{F(S)}^0({\bf x},t)\exp\bigg[
   i\int\frac{d^2kd\varepsilon}{(2\pi)^3}G^0_{F(S)}(k,\varepsilon)
   V_{S,S}(k,\varepsilon) G^0_{F(S)}(k,\varepsilon)
    \left(e^{i({\bf k}\cdot{\bf x} - \varepsilon t)}-1\right)\bigg]
    \label{eq:GF}
\end{equation}
\end{widetext}
This expression has many similarities with the bosonization formulas usually obtained
in one dimension. In particular, the free-fermion pre-factor also arises there.
However, the exponential factor, which in one dimension yields an anomalous
dimension for the fermion operator, plays a very different role in dimensions higher
than one.

\subsection{Diagrammatic expansion for the bosonized theory}
\label{sec:diaseries} 

We will first show that the bosonized formula of Eq.\eqref{eq:GF} is consistent with
the perturbative results of Ref.[\onlinecite{Oganesyan01}].
To do that we will expand the exponential and Fourier transform to momentum space to
find its diagrammatic expansion. The result will be a series of
convolutions since in real space they are products. The first
order term is:
\begin{widetext}
\begin{equation}
  \delta G_F^{(1)}({\bf q},\omega) =
  i\int\frac{d^2kd\varepsilon}{(2\pi)^3}
  \bigg(G_{F(S)}^0({\bf q-k},\omega-\varepsilon)-
     G_{F(S)}^0({\bf q},\omega)\bigg)
  \left[G_{F(S)}^0({\bf k},\varepsilon)\right]^2V_{S,S}(k,\varepsilon)
\end{equation}
\end{widetext}
which does not look like it obeys the Feynman rules for the
perturbation theory of non-relativistic fermions. However, we may
utilize the following identity
\begin{multline}
  \big(G_{F(S)}^0({\bf q-k},\omega-\varepsilon)-
     G_{F(S)}^0({\bf q},\omega)\big)G_{F(S)}^0({\bf k},\varepsilon)\\ =
  G_{F(S)}^0({\bf q-k},\omega-\varepsilon)G_{F(S)}^0({\bf q},\omega)
\end{multline}
and use it again through
\begin{multline}
  G_{F(S)}^0({\bf q-k},\omega-\varepsilon) =\\
  G_{F(S)}^0({\bf q-k},\omega-\varepsilon) -G_{F(S)}^0({\bf q},\omega) +
  G_{F(S)}^0({\bf q},\omega)
  \\
\end{multline}
To obtain
\begin{multline}\label{eq:rpa-firstorder}
  \delta G_F^{(1)}({\bf q},\omega) =
  i\left[G_{F(S)}^0(q,\omega)\right]^2 \times \\
  \int\frac{d^2kd\varepsilon}{(2\pi)^3}
  \bigg(G_{F(S)}^0({\bf q-k},\omega-\varepsilon)+
     G_{F(S)}^0({\bf k},\varepsilon)\bigg)
  V_{S,S}(k,\varepsilon)
  \\
\end{multline}
The second term is actually the shift in the chemical
potential, $\Sigma(k_F,0)$, and is zero by the effective particle-hole symmetry
of this theory (with a linearized fermion dispersion). We therefore obtain,
\begin{multline}
  \delta G_F^{(1)}({\bf q},\omega) =
  i\left[G_{F(S)}^0(q,\omega)\right]^2\times\\
  \int\frac{d^2kd\varepsilon}{(2\pi)^3}
  G_{F(S)}^0({\bf q-k},\omega-\varepsilon)
  V_{S,S}(k,\varepsilon)
\end{multline}
which is the correct result since, as usual, the Hartree term
vanishes.

\begin{figure}[!t]
  \centering
\includegraphics[width=0.45\textwidth]{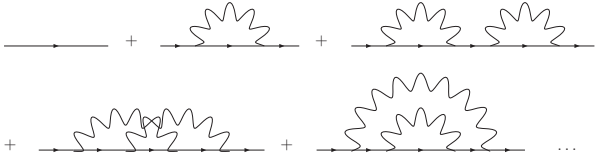}
 \caption{Bosonization's Feynman diagram series.}
  \label{fig:rpadiag}
\end{figure}

Using the same tricks and assumptions, the second order
contribution can also be worked out (though it is much more work).
The series to second order is shown in Fig. \ref{fig:rpadiag},
please keep in mind that the interaction, $V_{S,S}$ is the full
bubble summed RPA interaction. Higher dimensional bosonization
therefore keeps all diagrams in perturbation theory that contain
up to simple bubbles while neglecting more complicated bubbles as
is usual in the RPA (in the Landau theory of the Fermi liquid,
these irrelevant operators contribute subdominant potentially 
non-analytic temperature and frequency dependent terms
to physical quantities\cite{chubukov03}).

Some time ago,  Kopietz and Castilla\cite{kopietz96,kopietz97} used a somewhat different (and in principle equivalent) form of bosonization and discussed the effects of a quadratic term in the fermion energy dispersion. However, instead of using an operator identity (and thus not exploiting the non-perturbative character of bosonization) they chose to make contact with perturbation theory, and proceeded to propose a modified form for the fermion propagator directly. It gave an exponential factor similar to the one appearing here except the fermion Green functions that appear in the exponential include the quadratic terms in ther energy dispersion. However, these Green functions don't satisfy the criterion $\Sigma(k_F,0)=0$ so that in addition to the exponential, they include a pre-exponential factor that is necessary, for example, to cancel the second term of eq. \eqref{eq:rpa-firstorder}.  As a result, order-by-order in $V_{S,S}$, one needs to keep precisely the right pre-exponential factor to cancel the additional terms. A calculation with their method can therefore only be carried out to a small finite order, and one can no longer think of the exponential factor as separate from the pre-exponential factor.  In contrast, we have seen here that the bosonized expressions, when treated consistently, yield exact results which agree with those of perturbation theory order by order, albeit only in the low energy limit, including the singular behavior.

\subsection{Perturbative results}
\label{sec:pertresults}

Before computing the full non-perturbative form of the
fermion Green function, let us verify that our bosonized theory,
reproduces the
perturbative results of Oganesyan, Kivelson and Fradkin\cite {Oganesyan01} 
near the nematic QCP. We shall be interested,
therefore, in the integral:
\begin{equation}
  \Sigma^{(1)}_S({\bf q},\omega) = 
  i\int_{{\mathcal P}_S}\frac{d^2kd\varepsilon}{(2\pi)^3}
  G_{F(S)}^0({\bf q-k},\omega-\varepsilon)
  V_{S,S}(k,\varepsilon)\label{eq:sigma}
\end{equation}
Here $V_{ss}(k,\varepsilon)$ is the effective interaction mediated by the
collective modes, {\it c.f.\/} Eq.\eqref{eq:V2+}, Eq.\eqref{eq:V2-} and 
Eq.\eqref{eq:VST}. 
The main
contribution to the self-energy is due to the overdamped Goldstone
modes, as noted in Ref.[\onlinecite{Oganesyan01}]. 
Thus
we take a generic 
interaction of the form
\begin{equation}\label{eq:Vss}
  V_{S,S}(k,\varepsilon) = \frac{1/N(0)}{\delta(k)-K_0(s)}
\end{equation}
noting that $\cos^2 (2(\theta_S-\phi))$, which should appear as a
coefficient in the nematic case, only introduces irrelevant
contributions to the integral and is therefore left out of this
analysis. Here, at the QCP, $\delta(k)=-\kappa k^2/2$ but as a
check we may characterize the Fermi liquid phase by letting
$\delta(k)=\delta$, a constant.

Before computing $\Sigma^{(1)}_S({\bf q},\omega)$, we should note
that this expression gives the clearest definition of the patch.
Given the ultraviolet cutoff $k_F-\lambda/2 < |{\bf k}|<k_F
+\lambda/2$, the patch is defined so that in comparison to the
original
theory
\begin{equation}
  \Sigma^{(1)}({\bf k}={\bf k}_S+{\bf q},\omega) \approx
  \Sigma^{(1)}_S({\bf q},\omega),\quad {\bf q}\in {\mathcal P}_S
\end{equation}
and therefore the patch width is of order
\begin{equation}\label{eq:Lambda}
  \Lambda \sim \sqrt{k_F\lambda}
\end{equation}
dictated by the curvature of the Fermi surface alluded to earlier.

Now, to simplify our calculation, let us focus on the fermion lifetime 
near the Fermi surface:
\begin{equation}
  2\Gamma_S({\bf q},\omega) = 
  -2\sign(\omega){\mathcal Im}\Sigma_S({\bf q},\omega)
\end{equation}
This can most easily be expressed in terms of the spectral
function, $B_2({\bf q},s)$, derived from the imaginary part of
$V_{S,S}(q,s)$ as in Eq.\eqref{eq:spect} and Eq.\eqref{eq:VST} (after setting the
angular factors to 1)
\begin{multline}
 2\; \Gamma_S({\bf q},\omega)=\\
 \int_{-\Lambda/2}^{\Lambda/2}\!\frac{dk_t}{2\pi}
  \int^{q_n}_{q_n-|\omega|/v_F}\!\frac{dk_n}{2\pi}\;
  B_2\bigg(k,\displaystyle{\frac{\omega - v_F(q_n-k_n)}{kv_F}}\bigg)
  \\
\end{multline}
where $q_n$=${\bf v}_S\cdot{\bf q}/v_F$. Here we note that this integral is dominated
by the contribution of the overdamped mode, which enters in $B_2^+(q,s)$. (The other
contributions, associated with the propagating collective modes, only yield regular
dependences in the fermion frequency.)
Setting $q_n\to0$ and looking 
at the limit $\omega\to 0$ we obtain
\begin{equation}
 \Gamma_S(\omega)=
 \frac{1}{2\pi^2}\int_0^{\omega/v_F}dk_n\int_0^{\Lambda}dk_t
  B_2^+(\sqrt{k_n^2+k_t^2},\frac{k_n}{k_t})
\end{equation}
using normal and tangential coordinates. From here, we first integrate 
over the tangential momenta, look at small $q_n$ and obtain the long 
wavelength behavior. We find that in the Fermi Liquid phase 
($\delta(k)=\text{const}$)
\begin{equation}
  \Gamma_S(\omega)\sim \frac{1}{N(0)}\bigg(\frac{\omega}{v_F}\bigg)^2
    |\log \omega/\Lambda v_F|
\end{equation}
while at the nematic QCP, using $\delta=-\kappa k^2/2$:
\begin{equation}\label{eq:gammaqcp}
  \Gamma_S(\omega)\sim \frac{1}{N(0)}
      \bigg(\frac{\omega}{\kappa v_F}\bigg)^{2/3}
\end{equation}
In the Fermi liquid, the energy of the quasiparticle is
proportional to $\omega$ and we classify them as long lived.
However, at the nematic QCP, one can show via Kramers-Kronig that
the real part of the self energy also goes like $|\omega|^{2/3}$.
Hence at the nematic QCP, the life time of a quasi particle is not
well defined and to try to understand it as a perturbation about a
free theory of long lived quasi particles is meaningless. 

Thus, the bosonized theory reproduces the results found earlier on
perturbatively by Oganesyan and coworkers\cite{Oganesyan01} (see also
Ref.[\onlinecite{Metzner03,Metzner05}]). It
should be noted that the $\omega^{2/3}$ law was also 
found to appear in the perturbative calculation of the fermion self-energy
in a model of holes
interacting via a forward scattering $U(1)$ gauge interaction with a similar
form to \eqref{eq:Vss} in the context of high temperature
superconductors\cite{Lee89,Altshuler95}, and in the perturbative treatment of
the quantum critical point in a ferromagnetic metal.
\cite{Chubukov05}

\subsection{Non-perturbative results}
\label{sec:non-perturbative}

We now return to the non-perturbative bosonized expression for the fermion propagator
of Eq.\eqref{eq:GF}, which we will write as
\begin{equation}
  G_{F(S)}({\bf x},t) = Z_S({\bf x},t) G_{F(S)}^0({\bf x},t)
\end{equation}
In the {\em Fermi liquid phase} and at long distances and low frequencies
the factor $Z_S$ approaches a constant value,
$Z_S=Z_F<1$, {\it i.e.\/} the quasiparticle residue of the Fermi liquid state.  Our
goal here is to investigate the behavior of $Z_S({\bf x},t)$ near the nematic
QCP and in the nematic phase. However, given the complexity of the full analytic
expression, in this paper we will consider only on the equal-time, $t=0$,
behavior (sometimes called the ``one-particle density matrix'') and the 
equal-position, ${\bf x}=0$, dynamical correlation function, and only at zero
temperature. We will discuss its full behavior elsewhere.

\subsubsection{The equal-time fermion propagator}
\label{sec:equal-time}

As in the perturbative calculation, it is convenient to express
$\ln Z$ as an integral over the spectral functions $B_2^\pm(q,s)$. Once again, at the
{\em nematic QCP}, the
important contribution is due to the overdamped collective mode in 
$B_2^+(q,s)$, and we will neglect all other contributions. (This is an approximation which gives the long time behavior accurately. An expression valid for all times also includes the contribution of $B_2^-(q,s)$.)

Since we have set $t=0$ here, the result is quite simple
\begin{equation}
  \ln Z(x_n,0) = \!\int_0^{\lambda/2}\!\frac{dk_n}{2\pi} I_2(k_n)\big(\cos (k_n x_n)-1\big)
\end{equation}
where
\begin{equation}
\label{eq:I2}
  I_2(k_n) = \int_{-\Lambda/2}^{\Lambda/2}\frac{dk_t}{2\pi}\int_0^\infty\frac{d\omega}{2\pi}
  \frac{B_2^+(k,\omega/kv_F)}{\big(\omega+v_Fk_n\big)^2}
\end{equation}
We find, both from a numerical computation and from an analytic estimate, that
\begin{equation}
\label{eq:I2smallk}
  I_2(k_n) = \frac{1}{N(0)v_F}
    \frac{f_2\left(k_n\sqrt{\kappa}\right)}{\left(k_n\sqrt{\kappa}\right)^{4/3}}
\end{equation}
where $f_2(k_n\sqrt{\kappa})$ is a regular function of its argument. From this analysis, we
conclude after performing the final Fourier transformation:
\begin{equation}
\label{eq:nematicZ}
  Z_S\left(x_n,0\right) = C_P\exp\left\{-\frac{b}{N(0)v_F\sqrt{\kappa}}
    \left|\frac{x_n\!}{\sqrt{\kappa}}\right|^{1/3}\right\}
\end{equation}
valid for $|x_n|\gg\sqrt{\kappa}$. Here $b=0.0658$ and $C_P$ is a
constant factor resulting from subdominant terms in $I_2(k_n)$. 
This sharp decay of $Z_S(x_n,0)$, faster than any power law,
introduces a scale (similar to a correlation length arising from a
gap in the spectrum) and dominates over the Fermi liquid behavior
at low energies. From the above expression, the correlation length
is of order $\xi \sim \sqrt{\kappa}(k_F\sqrt{\kappa})^3$ which is
much longer than length of the interactions, $\sqrt{\kappa}$.

Let us also compute the behavior of $Z_S$ on the {\em nematic phase}
by focusing on the effect of the Goldstone modes. Recalling
our discussion of the order parameter theory, we replace
\eqref{eq:Vss} with
\begin{equation}
  V_{S,S}(k,\varepsilon) = \frac{\sin^2(2\theta_S)}{N(0)}
  \frac{1}{-\kappa k^2/2+i|s|\sin^2(2\phi)}
\end{equation}
which is the contribution from the Goldstone mode
($K_0(s)\approx-i|s|$). In the limit, $k_n\to0$, that is on the
Fermi surface, $\phi\to\theta_S+\pi/2$ and therefore, we simply
have
\begin{equation}
  V_{S,S}(k,\varepsilon) =
   \frac{1}{N(0)}\left(-\frac{\kappa k^2}{\sin^2(2\theta_S)}-K_0(s)\right)^{-1}
\end{equation}
so that the difference between this and the symmetric side is
simply that $\kappa\to\kappa/\sin^2(2\theta_S)$. Hence, we may
directly write down $Z_S$:
\begin{equation}
  Z_S\left(x_n,0\right) = C_{P(S)} \exp\left\{-
    \frac{b\left|\sin\left(2\theta_S\right)\right|^{4/3}}{N(0)v_F\sqrt{\kappa}}
    \left|\frac{x_n\!}{\sqrt{\kappa}}\right|^{1/3}\right\}
    \label{eq:cps}
\end{equation}
where $b$ is the same constant of Eq. \eqref{eq:nematicZ}. In Eq.\eqref{eq:cps} we have
not included the subdominant contributions which become the leading terms along the
symmetry-dictated directions, the ``nematic axes'', along which the angular factor 
vanishes. Along the nematic axes the  behavior of the equal-time correlation function 
has a more Fermi liquid like long distance behavior as shown by Oganesyan et. al., but due to the introduction of patchs, we cannot accurately capture this behavior here. 
   
Thus, we see that on the broken symmetry side, we have special points
at $\theta_S=n\pi/2$ where the Goldstone mode weakens and subdominant behavior takes
over. At these points, $Z_S=C_{P(S)} = Z_F<1$ and the
quasiparticles become long lived. It is interesting to note that
the angular dependence shown here with a power of $4/3$ is similar
to that of the perturbative calculation using the same
transformation of $\kappa\to\kappa/\sin^2(2\theta_S)$ on Eq.
\eqref{eq:gammaqcp} and it agrees with the results of Ref.[\onlinecite{Oganesyan01}].

\subsubsection{The fermion residue}
\label{sec:residue}
One simple calculation we can do with the above result for $Z_S(x_n,0)$ is the fermion residue following Migdal\cite{Migdal57}.
\begin{align}
	Z_q &= n({\bf k}_F -{\bf q}) - n({\bf k}_F+{\bf q})\\
		&= \int \frac{d\omega}{2\pi}\bigg(G_F({\bf k}_F - {\bf q},\omega) - 
					G_F({\bf k}_F + {\bf q})\bigg)e^{i\epsilon\omega}
\end{align} 
where the exponential factor tells us to close the contour in the upper half plane. This expression may be written in terms of the real space, real time fermion Green function, at time $t=-\epsilon$. Inserting our expression for $G_F$ within Bosonization and neglecting interpatch scattering (which should produce analytic in $q/\Lambda$ contributions here), we obtain:
\begin{equation}
	Z_{q_n} = \frac{2}{\pi} \int_0^{\infty} dx_n \frac{\sin q_n x_n}{x_n} Z_S(x_n,0) 
\end{equation}
\emph{In the Fermi liquid phase}, we find $Z_S(x_n,0) = Z < 1$ in the long distance limit and this leads directly to $Z_{q_n} = Z$ when $q_n\to0$ as expected. 

\emph{At the nematic QCP} and into the nematic phase away from the nodal points discussed earlier, $Z_S(x_n,0)$ is short ranged and we may expand to leading order in $q_n$ and obtain
\begin{equation}
	Z_{q_n} \approx \frac{12}{\pi}q_n\xi
\end{equation}
with the correlation length $\xi = \sqrt{\kappa}(\sqrt{\kappa}k_F/2\pi b)^3$. Thus the fermion residue vanishes linearly similar to how it would in a Fermi liquid at finite temperature where temperature makes the correlations short ranged. It should be noted, however, that because $Z_s(x_n,0)$ decays slower then $e^{-\alpha x_n}$ in the long distance limit, this series expansion that we have used is poorly defined at higher order with the coefficient of $q_n^{2j+1}$ growing so rapidly that the Taylor expansion has zero radius of convergence in the complex-$q_n$ plane. Thus $Z_{q_n}$ is not analytic in $q_n$ and the Fermi surface may still be defined as a singular point in $n({\bf k})$. The same behavior occurs {\em in the nematic phase} for generic momenta, except along the directions of the nematic principal axes where a finite residue is obtained.

\subsubsection{The fermion auto-correlation function}
\label{sec:autocorrelation}

This case turns out to be more complicated than the equal-time
expression and we present a full analysis in Appendix 
\ref{ap:equalt}. By a simple integration-by-parts, we found that
the double pole of Eq. \eqref{eq:I2} may be reduced to a single
pole and that the integrals involved were less singular by a full
power (diverging like $1/k_n^{1/3}$ unlike Eq.
\eqref{eq:I2smallk}) but with logarithmic corrections. From that
analysis, we found the general form of $Z_S$ to be
\begin{equation}
\label{eq:zst}
  Z_S(0,t) = C_P \exp\left\{-A(1\!-\!i\sqrt{3})
  \frac{\ln\left(v_F t/\sqrt{\kappa}\right)}{
            \left(v_F t/\sqrt{\kappa}\right)^{2/3}}\right\}
\end{equation}
In contrast with our result for the equal-time correlation function,
Eq.\eqref{eq:zst} approaches a constant at long times. However, it decays 
to that
constant much more slowly than in a Fermi liquid where we would
expect the exponent $2/3$ to become $2$. As a result, the non-perturbative 
effects are 
here less important and, consequently, this time-dependence
appears to exhibit the same power-law behavior as the life time
calculated perturbatively in Sec. \ref{sec:pertresults}.

The coefficient of the exponential was found to be ~~
$A\propto1/(k_F\lambda\kappa)$ and deserves some attention. In a
Fermi liquid, we would find $A\propto \lambda/k_F\ll1$ and the time
dependence would present a small correction from the Fermi-liquid
behavior. However, $A$ need not be small if
$1/\kappa \gg k_F\lambda \sim \Lambda^2$,
where $\Lambda$ is the patch width ({\it c. f.\/} Eq. \eqref{eq:Lambda}). Due to the emergence of $\Lambda$ here,
this limit occurs precisely where the Fermi-surface curvature
begins to matter, and where the interaction length scale is still
quite small ($k_F^{-1}\ll\sqrt{\kappa}\ll\Lambda^{-1}$).

In the nematic phase, again letting $\kappa\to\kappa/\sin^22\theta_S$, we find
\begin{multline}
\label{eq:Znemeqtime}
 Z_S(0,t) = \\C_{P(S)} 
 \exp\left\{-A(1\!-\!i\sqrt{3})\left|\sin2\theta_S\right|^{4/3}
            \frac{\ln\left(\frac{v_F t}{\sqrt{\kappa}}|\sin2\theta_S|\right)}{
            \left(v_F t/\sqrt{\kappa}\right)^{2/3}}\right\}
\end{multline}
so that the angular dependence is similar to that of the equal-time behavior.

\subsubsection{The one-particle density of states}
\label{sec:1dos}

We close this section with  an application of these results to the
calculation of the one-particle density of states (DOS) in both
the Fermi liquid and the nematic phases, and at the
nematic (Pomeranchuk) quantum critical point.

The one-particle DOS is defined by the standard expression
\begin{equation}
  N^*(\omega) = - \sign(\omega)\; \frac{1}{\pi} 
  \textrm{ Im}\; G_F(x,x; \omega)
\end{equation}

{\em In Fermi liquid phase}, using $Z_S(0,t) = C_F$, a constant dependent upon the Fermi liquid parameters that goes to 1 for the non-interacting case, as expected we find
\begin{align}
  G_F(x,x,\omega) &= \sum_S \frac{1}{N}  
  C_F\int dt G^0_{F(S)}(0,t)e^{i\omega t} \nonumber \\
    &= C_F\int dt \frac{N(0)v_F}{-v_F t + i a \sign (t)} e^{i\omega t} 
    \nonumber \\
    &= -i \pi \sign(\omega)C_F N(0)
\end{align}
so that here, $N^*(\omega) = C_F N(0)$ as expected. Note: $C_F$ is not the fermion residue and may be greater than one.

{\em At the nematic QCP}, we have instead
\begin{widetext}
\begin{equation}
\label{eq:GF-nematic-qcp}
  G_F(x,x,\omega) = \sum_S\frac{1}{N} \int_{-\infty}^\infty dt \; C_P\; e^{i\omega t}
    \exp\left\{-A(1\!-\!i\sqrt{3})\frac{\ln\left(v_F t/\sqrt{\kappa}\right)}{
    \left(v_F t/\sqrt{\kappa}\right)^{2/3}}\right\}
    G^0_{F(S)}(0,t)
\end{equation}
\end{widetext}
so that $N^*(\omega) = C_P N(0) I(\bar \omega)$ with
\begin{multline}
\label{eq:Iomega0}
  I(\omega) = \frac{2}{\pi}{\mathcal Re}\int_0^{\infty} \frac{du}{u}
   \sin\left(\bar\omega u\right)
   \exp\left\{-A(1\!-\!i\sqrt{3})\frac{\ln u}{u^{2/3}}\right\}
   \\
\end{multline}
with $\bar\omega = \sqrt{\kappa}\omega/v_F$, valid for $\omega \ll
v_F \min \{\kappa^{-1/2},\lambda\}$. Notice that we have used throughout these expressions only the long time limit of the exponential factor. At shorter times the behavior of the exponential should be dominated by high energy effects which are insensitive to whether the system is in a Fermi liquid, a quantum critical point or in a nematic phase. Thus, the time integrals have an implicit short distance cutoff (which we have denoted by ``$0$''). In any event, we are only interested in the low frequency behavior which is dominated by the long time part of the integration range.

By inspection we see that the 
exponential factor
approaches unity (quite rapidly) for large $u\gg 1$. 
Thus, the main effect of this factor
is a correction to $I(\omega)$ away from its value 
at zero frequency, {\it i.e.\/}, 
$I(0)=1$. The leading finite frequency behavior, as 
$\omega\to0$, is obtained by
expanding the exponential factor in Eq.\eqref{eq:Iomega0}:
\begin{eqnarray}
   I(\omega) &&= 1 - \frac{2A}{\pi}\int_0^\infty \frac{du}{u}
      \frac{\sin\bar\omega u}{u^{2/3}}\ln u +\ldots \nonumber \\
    &&=1+A\; \frac{3\sqrt{3}}{2\pi}\Gamma(1/3)
    \left(\frac{\sqrt{\kappa}\omega}{v_F}\right)^{2/3} \; 
    \ln \left(\frac{\sqrt{\kappa}\omega}{v_F}\right)+\ldots
   \nonumber \\
   &&
   \label{eq:Iomega}
\end{eqnarray}
where the last line is accurate for $\sqrt{\kappa}\omega/v_F\ll 0.1$. 
The ellipsis in Eq.\eqref{eq:Iomega} represents subdominant contributions at 
low frequencies, which vanish faster than $\omega^{2/3} \ln \omega$ as $\omega \to 0$.
As a result, the $\omega^{2/3}$ behavior of the
inverse life time $\Gamma(\omega)$ calculated perturbatively,
appears here as a cusp in the DOS (with a logarithmic correction).
Unlike the lifetime, though,
here $A$ depends on the product of the patch width cutoff
$\sqrt{k_F\lambda}\sim\Lambda$ and $\sqrt{\kappa}$ while
$\Gamma(\omega)$ depends only on the product of $k_F$ and
$\sqrt{\kappa}$. 

Thus we find that the low frequency one-particle density of states has the form \begin{equation}
\label{eq:Nomega}
N^*(\omega)=N^*(0)+B\; \omega^{2/3} \ln \omega +\ldots
\end{equation}
 where $N^*(0)=C_P N(0)> N(0)$,   since we found the constant $C_P>1$ (see Eq.\eqref{eq:CP}), and $B=A C_P N(0) \frac{3\sqrt{3}}{2\pi} \Gamma(1/3)$ (see Eq.\eqref{eq:A}). Hence,  the zero frequency value of the density of states is {\em larger} than the Fermi liquid value. As the frequency increases, the one-particle density of states decreases from its zero frequency value according to the $\omega^{2/3} \ln \omega$ correction term. This is a cusp singularity at $\omega=0$. It is important to stress that we obtained these expressions upon  expanding  the exponential factor in the auto-correlation function. This is consistent since this factor asymptotically (and rather rapidly) approaches $1$ at long times. This also implies that, in this regime, our results should be consistent with the behavior of the fermion Green function found in perturbation theory around Hartree-Fock/RPA\cite{Oganesyan01,Metzner03,Metzner05,Chubukov05}.

{\em In the nematic phase}, an angular average of Eq.
\eqref{eq:Znemeqtime} enters our expression for $N^*(\omega)$. In
the long time limit, when the argument of the exponential is much
less than one, we expect that for low frequencies
\begin{equation}
  A \to \left\langle\big|\sin2\theta_S\big|^{4/3}\right\rangle A\approx0.58A
\end{equation}
Hence, in the nematic phase, the Goldstone modes continue the
critical behavior of this function in only a mildly weaker form.

In summary, in this section we used bosonization to compute the 
non-perturbative behavior
of the fermion propagator using the bosonized form of the fermion operator. 
We first
checked that the non-Fermi liquid behavior of the nematic QCP and in the
nematic phase, which were obtained earlier using conventional diagrammatic 
(perturbative)
methods,\cite{Oganesyan01,Metzner03} is
recovered here upon expanding the bosonized expression to leading order in $V_{SS}$, {\it i.e.\/} a single boson exchange. 
However, upon a closer
examination of the full bosonized result we found that the 
equal-time fermion correlation
function has a  much more singular behavior that could have 
been predicted in perturbation
theory. In contrast, the fermion auto-correlation function 
(and hence the one -particle
density of states) is seemingly consistent with the perturbative 
analysis of the quantum
critical behavior.  

We note here that Chubukov\cite{Chubukov05},
has analyzed the quantum critical behavior of ferromagnetic 
Fermi liquids and 
claims that the $\omega^{2/3}$ behavior, found at lowest order, 
persists to all orders in
perturbation theory.  The results of this section for the fermion auto-correlation function appear to agree with those of Chubukov and coworkers. However our results for the equal-time fermion correlator apparently disagree with these results.
Clearly, a more detailed analysis of the bosonized expression for this 
propagator is
warranted. We will discuss this problem in  a separate publication.
 
\section{Conclusion}
\label{sec:conc}

In this paper, we have utilized the method of high dimensional
bosonization to  study non-perturbatively the quantum phase transition from a
Landau Fermi liquid state to a nematic phase, a nematic (Pomeranchuk)
instability. For this purpose, we have constructed an order
parameter theory from the boson theory by integrating out
non-critical modes and verified that this boson theory is
equivalent to RPA. We then turned to studying the bosonization
form of the fermion propagator and found its diagrammatic expansion
proving the correctness and clarifying the arguments leading up to
that expression. This diagrammatic expansion keeps all diagrams up
to the simple bubble in the spirit of RPA as applied to the
density-density propagator and shows, in specific, that
bosonization goes beyond the self-consistent Born approximation to
include vertex corrections. We then found explicitly that
bosonization reproduces the results of Hartree-Fock with an RPA
interaction by showing that the lifetime computed in this limit
also has an $|\omega|^{2/3}$ dependence as originally found for
the case of the nematic QCP by Oganesyan and coworkers \cite{Oganesyan01}. 

Lastly, we calculated the fermion propagator non-perturbatively and found the
dramatic effect of the overdamped critical mode that induces short
ranged spatial correlations that decay nearly exponentially $\frac{1}{|x|}
e^{- {\rm const.}\; |x|^{1/3}}$, while the auto-correlation function exhibits a
milder non-Fermi liquid behavior of the form $\frac{1}{|t|}\; \exp(-{\rm
const}.\; |t|^{-{2/3}})$. 
From this short ranged behavior of the equal time Green function, we verify that the fermion residue vanishes at the critical point and into the nematic phase except at four special points. We also calculated the one-particle (fermion) density of states
$N^*(\omega)$. We found the low frequency behavior $N^*(\omega)=N^*(0)+ B\; \omega^{2/3} \ln \omega$ (with $N^*(0)>N(0)$) both at the quantum critical point and into the nematic phase. 
Thus, the fermion propagator exhibits unexpected behaviors which could not have 
been anticipated by the existing perturbative
results\cite{Oganesyan01,Metzner03,Chubukov04,Chubukov05,Nilsson05}. In a separate publication we will
present a more detailed analysis of the fermion spectral function in both 
phases and at finite temperature.

Two recent papers, one by Yang \cite{Yang05} and another by Nilsson and 
Castro
Neto\cite{Nilsson05}, also study Pomeranchuk nematic instabilities in
two-dimensional Fermi
systems. Yang also derives an order parameter
theory within high dimensional bosonization. However, contrary to our results, 
concludes that
the critical mode has $z=2$ and is undamped. While we agree that a
$z=2$ propagating mode does exist at the critical point, we find that the
spectral function is completely dominated by the
overdamped $z=3$ mode.
This effect is due to Landau damping, a consequence of the curvature of the Fermi
surface, and dominates the low energy behavior of the theory at the critical
point.  The reason for this disagreement is that in
Ref.[\onlinecite{Yang05}] the effects of Landau damping are ignored. We find
that these effects are crucial.

On the other hand, Nilsson and Castro Neto\cite{Nilsson05} approach the 
nematic quantum phase transition within
the more traditional methods found in the Fermi liquid theory
literature. They first find an order parameter theory by
constructing a path integral, whose classical equations of motion
give the collisionless Boltzmann equation as in high dimensional
bosonization, and integrating out the non-critical modes. Their
results, however, agree with ours in all essential details,
including the existence of a $z=3$ mode. They then calculate the
fermion lifetime using the Bethe-Salpeter equations and
Fermi's Golden rule finding that $\tau^{-1}\propto
\epsilon^{2/3}$, as in \eqref{eq:gammaqcp}, and agree with us, in the
perturbative regime, in
concluding that this represents both a breakdown of Fermi liquid
theory and perturbation theory.

\begin{acknowledgments}
We would like to thank Steve Kivelson, Vadim Oganesyan, Antonio Castro Neto and Andrey Chubukov for many insightful discussions and comments.
This work was supported, in part, by the National Science Foundation (USA) through the 
 grants No.
DMR-01-32990 and DMR 04-42537 at the University of Illinois
(ML,EF,VF), by Fundaci\'on Antorchas (Argentina) (VF) and by the Conselho
Nacional de Desenvolvimento Cient{\'\i}fico e Tecnol\'ogico
(CNPq-Brazil) (DGB,LO), the Funda\c{c}ao de Amparo \`a Pesquisa do Estado
do Rio de Janeiro (FAPERJ, Brazil) (DGB,LO), CAPES (DGB, LO), and SR2-UERJ (Brazil) (DGB).
D. Barci is a Regular Associate of the ``Abdus Salam International Centre for Theoretical Physics, ICTP'', Trieste, Italy.
\end{acknowledgments}

\appendix
\section{Summary of Bosonization in $D$-dimensional Fermi Systems}
\label{ap:bosonintro} 

Consider the Fermi liquid theory of spinless
fermions interacting via a short but perhaps finite range forward
scattering interaction living in a translationally invariant
D-dimensional world. This is a low energy theory and as such,
following Landau we shall linearize the energy dispersion near the
Fermi surface, though corrections to this may be considered when
necessary. To this end, let us build a construction in which we
linearize within $N$ equally sized patches approximating the Fermi
surface. For this construction to be reasonable, our end result
should be relatively insensitive to the details belonging to this
partitioning. Keep in mind, however, that a remarkable property of
Fermi liquids is that they only require a few of the lowest
angular momentum Fermi liquid parameters to understand a wide
range of phenomena.

The number of patches, $N$, approximating the Fermi surface will
naturally be inversely proportional to its curvature. As such, if
the density, $n\to\infty$ then $N\to\infty$ and the construction
becomes exact. An exact solution to leading order in $N$ of this
linearized theory is therefore equivalent to the asymptotic low energy limit as dictated by the renormalization group. More specifically, the number of patches $N$  and the patch width $\Lambda\sim \sqrt{k_F \lambda}$, where $\lambda$ is the energy cutoff, must be related by the condition $2\pi k_F=N \Lambda$, required for the Fermi system to a have a finite density of states and curvature (see below). Consequently, the number of patches $N$ must scale as $N\sim 2\pi \sqrt{k_F/\lambda}$. Clearly $N\to \infty$ in the infrared limit $\lambda \to 0$. Many
of the results of this paper can be understood in light of this
basic reasoning.

Under our construction, the fermion annihilation operator becomes:
\begin{equation}
\begin{split}
  \hat c({\bf x})
  &= \sum_{\bf k} \hat c_{\bf k} \frac{e^{i{\bf k}\cdot{\bf x}}}{\sqrt{L^D}}\\
  &= \sum_S\sum_{{\bf q}\in{\mathcal P}_S} \hat c_{{\bf k}_S+{\bf q}}
     \frac{e^{i({\bf k}_S+{\bf q})\cdot{\bf x}}}{\sqrt{L^D}}\\
  &= \sum_S\hat c_S({\bf x})\frac{e^{i{\bf k}_S\cdot{\bf x}}}{\sqrt{N}}
\end{split}
\end{equation}
where $S$ labels the patch, ${\mathcal P}_S$ is the volume in
${\bf k}$-space around the point ${\bf k}_S$ and the new fermion
operators, $\psi_S({\bf x})$, obey the canonical commutation
relations
\begin{equation}
  \big\{c_S({\bf x}), c^\dagger_T({\bf x'})\big\} = 
  \delta_{S,T}\delta^D({\bf x-x'})
\end{equation}
It is also important to note the Fourier transform normalizations 
within the construction:
\begin{equation}
\begin{split}
  &\delta^D({\bf x-x'}) = \delta^{D-1}_t({\bf x_t-x_t'})\delta(x_n-x_n')\\
  &= \bigg(N\!\sum_{|{\bf q_t}|<\frac{\Lambda}{2}}
        \frac{e^{i{\bf q_t}\cdot({\bf x_t-x_t'})}}{L^{(D-1)}}\bigg)
    \bigg(\sum_{|{\bf q_n}|<\frac{\lambda}{2}}\frac{e^{iq_n(x_n-x_n')}}{L}\bigg)
\end{split}
\end{equation}
where we have introduced $\Lambda$ to characterize the tangential
width of the patch ($N\Lambda$ is the area of the Fermi surface)
and $\lambda$ as an ultraviolet cutoff about the Fermi surface.

Now, in terms of our fermion operators, the linearized Hamiltonian is
\begin{equation}
  :\!\hat H\!:\ = \sum_S\int d^{\bf x}\ \bigg({\mathcal H}^0_S({\bf x}) +
                  {\mathcal H}^{\text{Int}}_S({\bf x})\bigg)
\end{equation}
where the free Hamiltonian density is
\begin{equation}
  {\mathcal H}_0({\bf x}) = \frac{\hbar}{2i}{\bf v}_S\cdot\bigg(
    :\!\big({\bf\nabla}\hat c^\dagger_S({\bf x})\big)\hat c_S({\bf x})
    -\hat c^\dagger_S({\bf x}){\bf \nabla}\hat c_S({\bf x})\!:\bigg)
    \label{eq:H0}
\end{equation}
and the forward scattering interactions are described by
\begin{equation}
  {\mathcal H}^{\text{Int}}_S({\bf x}) = \sum_T\int d{\bf x}' F_{S-T}({\bf x-x'})
    \delta \hat n_S({\bf x})\delta\hat n_T({\bf x'})
\end{equation}
Here, the density fluctuations are defined by
\begin{equation}
  \delta \hat n_S({\bf x}) \equiv\ :\!\hat n_S({\bf x})\!:\ = \
    :\!\hat c_S^\dagger({\bf x})\hat c_S({\bf x})\!:
\end{equation}
and throughout this description we have been using the usual normal 
ordering procedure for any operator ${\mathcal O}$:
\begin{equation}
  :\!{\mathcal O}\!: \big|G\rangle = 0 \to
  {\mathcal O}\ =\ :\!{\mathcal O}\!: - 
  \langle G \big| {\mathcal O} \big|G\rangle
\end{equation}
where we take the filled Fermi sea as our ground state:
\begin{equation}
  \big|G\rangle = 
  \prod_S\prod_{\{{\bf q}\in{\mathcal P}_S|{\bf v}_S\cdot{\bf q}<0\}}
     \hat c_{{\bf q},S}^\dagger\big|0\rangle
\end{equation}

It was shown by Haldane\cite{Haldane92,Haldane94}, Castro Neto and
Fradkin\cite{CastroNeto93,CastroNeto94,CastroNeto95} and Houghton
and Marston\cite{Houghton93,Houghton00} that this Hamiltonian can
be entirely described in terms of the electron density operators,
$\delta \hat n_S({\bf x})$ in the high density limit and it
is quadratic in these operators.  A fermion operator may then be
constructed following well known 1D bosonization techniques and so
the theory can be solved exactly. (This
represented a major step forward since the introduction of RPA by
Bohm and Pines\cite{Pines66}.) Here we shall outline the proof of this solution,
but with the traditional approach of point-splitting regularization, commonly used
in the one-dimensional case (see for example\cite{Affleck86}).

The expectation value of the density operator $\hat n_S({\bf x})$
in the ground state of the Fermi sea is clearly divergent if we
send the density of fermions to infinity. As a result, in the high
density limit we are interested in, it is poorly defined. To
control this divergence, we introduce the point-split operator:
\begin{equation}
\begin{split}
  \hat n_S^{\epsilon}({\bf x}) &\equiv
    \hat c_S^\dagger({\bf x+\epsilon/2})\hat c_S({\bf x-\epsilon/2})\\
  &=\ :\!\hat n_S^{\epsilon}({\bf x})\!: -
    \langle G\big|\hat n_S^{\epsilon}({\bf x})\big|G\rangle
\end{split}
\end{equation}
Here, by short-distance it is meant a length scale short compared with the separation of all operators of interest but long compared with physical short length scales, {\it i.e.\/}
It should be noted that physically point-splitting can be thought of as a means of 
$|{\bf x}|\gg|\epsilon| \gg
\lambda^{-1}$.

The divergent part may be computed explicitly:
\begin{equation}
\begin{split}
 \langle G\big|\hat n_S^{\epsilon}({\bf x})\big|G\rangle &=
    \frac{N}{L^D}\sum_{{\bf q_t}}\sum_{q_n<0}e^{-i{\bf q}\cdot{\bf \epsilon}}\\
  &=\delta^{D-1}_t({\bf \epsilon_t})
    \int_{-\infty}^{0}\frac{dq_n}{2\pi}e^{-iq_n\epsilon_n + q_n/\lambda}\\
  &=\frac{i\delta^{D-1}_t({\bf \epsilon_t})}{2\pi(\epsilon_n+i\lambda^{-1})}
\end{split}
\end{equation}
where we have implemented the ultraviolet cutoff, $\lambda$, as a
soft cutoff $e^{-|q_n|/\lambda}$. This is a highly anisotropic
expression in ${\bf\epsilon}$. It vanishes if we first send
$\epsilon_n\to0$, looking along a tangential direction
(${|\bf\epsilon}_t|\gg\Lambda^{-1}\to0$) but diverges if we first
send $\epsilon_t\to0$ looking along
$\epsilon_n\gg\lambda^{-1}\to0$. Therefore, to capture the basic
physics of the density operator we must choose the latter limit as
the definition of the point split operator. Keeping in mind that
$\delta_t(0)=A_F/(2\pi)^{D-1}=2\pi\hbar v_FN(0)$, where $A_F$ is
the $(D-1)$-dimensional Fermi surface area and $N(0)$ is the
density of states at the Fermi surface, we obtain
\begin{equation}
 \hat n_S^{\epsilon_n}({\bf x}) = -\frac{iN(0)\hbar v_F}{\epsilon_n} +
    \delta\hat n_{S}({\bf x}) + 
    \frac{i\epsilon_n}{\hbar v_F}{\mathcal H}_S^0({\bf x})
    + \ldots\label{eq:pointsplit}
\end{equation}
to leading order in $\epsilon_n\gg\lambda^{-1}$ and where we kept
the expansion of $:\!\hat n_S^{\epsilon_n}({\bf x})\!:$ to first
order noticing the useful emergence of the free Hamiltonian
density operator.

Now that we have a controlled definition of the density operator, 
we proceed with computing its commutator:
\begin{multline}
  \big[\hat n^{\epsilon_n}_S({\bf x}), \hat n^{\epsilon_n'}_T({\bf x'})\big] =
  \frac{i N(0)\hbar v_F}{\epsilon_n+\epsilon_n'}
  \delta_{S,T}\delta_t^{D-1}({\bf x_t-x_t'})\\
  \times\bigg(\delta\big(x_n-x_n' - (\epsilon_n+\epsilon_n')/2\big) -
        \delta\big(x_n-x_n' + (\epsilon_n+\epsilon_n')/2\big)\bigg)
\end{multline}
Expanding both sides of this equation in powers of $\epsilon_n$ and 
equating like powers gives us the following result
\begin{align}
  \big[\delta\hat n_S({\bf x}), \delta\hat n_T({\bf x'})\big] &=
  -i\hbar N(0)
  \delta_{S,T}{\bf v}_S\cdot\nabla\delta^{D}({\bf x-x'})\label{eq:kacmoody}\\
  \big[\delta\hat n_S({\bf x}),{\mathcal H}_{T}^0({\bf x'})\big] &=
    -i\hbar{\bf v}_S\cdot\nabla\delta\hat n_S({\bf x})\delta^{D}({\bf x-x'})
\end{align}
Using these commutators, we may compute the equation of motion for the 
Heisenberg operator $\delta\hat n_S({\bf x},t)$:
\begin{multline}
  \partial_t\delta\hat n_S({\bf x},t) + 
  {\bf v}_S\cdot\nabla\delta\hat n_S({\bf x},t) +\\
  {\bf v}_S\cdot\nabla\sum_T\int d{\bf x}'F_{S-T}({\bf x-x'})
  \delta\hat n_T({\bf x'},t) = 0
\end{multline}
where $F_{S-T} = N(0)f_{S-T}$. This is the linearized
collisionless Boltzmann equation in operator form found by Castro
Neto and Fradkin in the context of a coherent state
formalism\cite{CastroNeto93}. We also notice through this
derivation that we may let
\begin{equation}
  {\mathcal H}_S^0({\bf x}) = \frac{1}{2N(0)}\delta\hat n_{S}^2({\bf x})
\end{equation}
and obtain exactly the same answer. Hence, the Hamiltonian may be
expressed entirely in terms of the density operator $\delta\hat
n_{S}({\bf x})$.

A natural consequence of \eqref{eq:kacmoody} is that the density
operator may be expressed in terms of a chiral boson field:
\begin{equation}
  \delta\hat n_{S}({\bf x}) = N(0){\bf v}_S\cdot\nabla\hat\varphi_{S}({\bf x})
\end{equation}
with a canonically conjugate momentum
\begin{equation}
  \hat\pi_{S}({\bf x}) = -N(0){\bf v}_S\cdot\nabla\hat\varphi_{S}({\bf x})
\end{equation}
These chiral bosons are a direct extension of the right / left
chiral bosons in the context of 1D bosonization. Following this
extension then, we may express the fermion operator as a vertex
operator
\begin{equation}\label{eq:psi}
  \hat\psi_S({\bf x}) = \eta_S({\bf x}_t)\sqrt{N(0)v_F\lambda}
      :\!e^{-i\hat\varphi_S({\bf x})/\hbar}\!:
\end{equation}
where $\eta_{S}({\bf x}_t)$ is a set of Klein factors responsible
for ensuring that $\hat\psi_S({\bf x})$ obey the proper
anti-commutation relations within the patch and on different
patches. The relation between this fermion operator and the
original $\hat c_{S}(x)$-operators will be made precise in Section
\ref{sec:diaseries} via direct comparison of the perturbation
series in $F_{S-T}({\bf x-x'})$ obtained in the bosonized theory.
$\hat\psi_S({\bf x})$, therefore, is equivalent to $\hat c_S({\bf
x})$ in the free case and in the interacting case, it is this
operator projected onto the high-density subspace. Hence, this is
an effective low energy theory of a dense Fermi system and
$\hat\psi^\dagger_S({\bf x})$ can be viewed as actually creating
Landau quasiparticles in the Fermi liquid phase.

The canonical structure of the bosonized theory also allows us to
directly write down a path integral formulation of the problem, including interactions described by a set of Landau parameters
(for a derivation using coherent states, see
Ref.[\onlinecite{CastroNeto93}]). The action for the bosonized theory has the general form
\begin{widetext}
\begin{equation}
  S = \frac{N(0)}{2}\sum_S\!\int d^2xdt
  \bigg[-\partial_t\varphi_S{\bf v}_S\cdot\nabla\varphi_S-
  \big({\bf v}_S\cdot\nabla\varphi_S\big)^2\bigg]
+
  \frac{N(0)}{2}\sum_{S,T}\int d^2xd^2x'dt 
    F_{S-T}({\bf x-x'})
    {\bf v}_S\cdot\nabla\varphi_S({\bf x}){\bf v}_T\cdot\nabla
    \varphi_T({\bf x'})
\end{equation}
\end{widetext}
This action is a  quadratic form in the Bose fields. Here we have not included higher order terms (such as those discussed in the body of the paper, in the context of the nematic instability). Such terms are generally present due to non-linearities in the fermion dispersion relation, as well as many-body effective 
interactions.\cite{Barci03} We have also not included ``vertex operators'' such as those associated with pairing (BCS) interactions.\cite{CastroNeto93,Houghton00}
We shall find it more convenient to work within this path-integral formulation
when constructing a theory of the nematic quantum critical
point.

Before leaving our discussion of high dimensional bosonization, we should make a final comment on its validity.  As discussed in the main body of the paper, the boson theory here completely recovers the random phase approximation (RPA) in the long wavelength limit. This should not be surprising since, in the asymptotic low energy limit,  both RPA and bosonization saturate the f-sum rule and are (formally) exact.  This is a well known established property of bosonization, extensively discussed in the literature since the 1970's. 

\section{Analysis of $V_2^+$ and $V_2^-$}
\label{ap:part}

Here we present the partial fraction expansion of the effective interactions $V_2^+$ and $V_2^-$ and their behavior near quantum criticality. 

\subsection{Partial fraction expansion of $V_2^+(q,s)$}

In terms of the function $K_0(s)$ we may write $V_2^+$ as
\begin{equation}
\label{eq:V2+-app}
\begin{split}
  V^{+}_2(q,s) &= \frac{1}{N(0)}\Bigg[\delta_2(q)-
    K_0(s)\bigg(1+\bigg(\tfrac{1-K_0(s)}{1+K_0(s)}\bigg)^2\bigg)\Bigg]^{-1}\\
               &= \frac{1}{2N(0)}\bigg[\frac{(1+x)^2}{
    x^3-\frac{\delta_2}{2}x^2+(1-\delta_2(q))x-\frac{\delta_2(q)}{2}}\bigg]
\end{split}
\end{equation}
where $x=K_0(s)$. The denominator is thus a cubic polynomial in
$K_0(s)$ and solving for the poles, $x=\delta_\beta^+(q)$, we find
\begin{multline}
  \delta_\beta^+(q) = \bigg\{\frac{1}{6}\big(\delta_2(q) +
          \frac{\delta_2(q)^2+12\delta_2(q)-12}{e^{in\pi/3}h(\delta_2(q))}\\
           + e^{in\pi/3}h(\delta_2(q))\big)\bigg|n\in\{-1,1,3\}\bigg\}
\end{multline}
where
\begin{equation}
 h(x) = (12\sqrt{12-36x+42x^2+3x^3}-x(x^2+18x+36))^{1/3}
\end{equation}
and we assign $\beta=\{a,b,c\}$ to each of these three poles. The
partial fraction expansion is
\begin{equation}
  V^+_2(q,s)=\frac{1}{N(0)}\sum_{\beta=a,b,c} 
  \frac{{\mathcal Z}^+_\beta(q)}{\delta^+_\beta(q)-K_0(s)}
\end{equation}
with residues
\begin{equation}
\begin{split}
{\mathcal Z}_a^+(q) &= \frac{1}{2}
    \frac{(1+\delta_a^+)^2}{(\delta_a^+-\delta_b^+)(\delta_a^+-\delta_c^+)}\\
{\mathcal Z}_b^+(q) &= \frac{1}{2}
    \frac{(1+\delta_b^+)^2}{(\delta_b^+-\delta_a^+)(\delta_b^+-\delta_c^+)}\\
{\mathcal Z}_c^+(q) &= \frac{1}{2}
    \frac{(1+\delta_c^+)^2}{(\delta_c^+-\delta_a^+)(\delta_c^+-\delta_b^+)}
\end{split}
\end{equation}
For $|\delta_2(q)|\ll 1$ these formulas simplify to:
\begin{align}
  \delta_a^+ &\approx i, &\delta_b^+&\approx -i, &\delta_c^+&\approx\delta_2/2\\
  {\mathcal Z}_a^+ &\approx i/2 , &{\mathcal Z}_b^+ &\approx -i/2,
  &{\mathcal Z}_c^+ &\approx 1/2
\end{align}
as a result we may write near the nematic QCP:
\begin{equation}
  V_2^+(q,s) \approx \frac{1}{2N(0)}\bigg[\frac{1}{\frac{\delta_2(q)}{2}-K_0(s)}
    - \frac{1/4}{s^2-1/2}\bigg]
\end{equation}

\subsection{Partial fraction expansion of $V_2^-(q,s)$}

We may write $V_2^-(q,s)$ as
\begin{multline}
  V^{-}_2(q,s) = \frac{1}{N(0)}
  \bigg[\delta_2(q)-
    K_0(s)\bigg(1-\bigg(\tfrac{1-K_0(s)}{1+K_0(s)}\bigg)^2\bigg)\bigg]^{-1}\\
               = \frac{1}{(4-\delta_2(q))N(0)}\bigg[\frac{1}{(1+x)^2}{
    x^2-2\frac{\delta_2(q)}{4-\delta_2}x-\frac{\delta_2}{4-\delta_2}}\bigg]\\
\end{multline}
The denominator is simply quadratic and we find poles at 
$x=\delta_\beta^-$ with
\begin{equation}
  \delta_a^-(q) = -\frac{\sqrt{\delta_2(q)}}{\sqrt{\delta_2(q)} + 2},\quad
  \delta_b^-(q) = -\frac{\sqrt{\delta_2(q)}}{\sqrt{\delta_2(q)} - 2}
\end{equation}
Hence, we can now expand in partial fractions to obtain
\begin{equation}
  V^-_2(q,s)=\frac{1}{N(0)}
  \bigg(\frac{{\mathcal Z}^-_a(q)}{\delta^-_a(q)-K_0(s)}+
  \frac{{\mathcal Z}^-_b(q)}{\delta^-_b(q)-K_0(s)}\bigg)
\end{equation}
with residues
\begin{equation}
{\mathcal Z}_a^- = \frac{1}{4-\delta_2}
\bigg[\frac{(1+\delta_a^-)^2}{\delta_a^--\delta_b^-}\bigg],\quad
{\mathcal Z}_b^- = \frac{1}{4-\delta_2}
\bigg[\frac{(1+\delta_b^-)^2}{\delta_b^--\delta_a^-}\bigg]
\end{equation}
Again, these formulas simplify for $|\delta_2|\ll1$:
\begin{align}
  \delta_a^-&\approx - \frac{\sqrt{\delta_2}}{2},  
  &\delta_b^-&\approx \frac{\sqrt{\delta_2}}{2}\\
  {\mathcal Z}_a^- &\approx -\frac{1}{4\sqrt{\delta_2}},
  &{\mathcal Z}_b^- &\approx \frac{1}{4\sqrt{\delta_2}}
\end{align}
Thus, near the Nematic QCP we can write:
\begin{equation}
  V_2^-(q,s) \approx \frac{1}{4N(0)}
  \bigg(\frac{1}{s^2+\frac{\delta_2(q)}{4}}\bigg)
\end{equation}

\section{Inclusion of $F_0(q)$}
\label{ap:F0}

Here we calculate $V_{S,T}$ for the case when both $F_0$ and $F_2$
are present and we let $F_2$ approach the Nematic QCP
($F_2\to-1$). We shall find, however, that the critical behavior
is utterly independent of $F_0$! Now, let
\begin{equation}
 F_{S-T}(q) = 
 \frac{2}{N}\bigg(F_0 + F_2(q)\cos\big(2(\theta_S-\theta_T)\big)\bigg)
\end{equation}
Returning to the $\sigma_{\ell}^\alpha$-theory, whose correlators
are the RPA interaction $V_{S,T}$, we find in this case
\begin{multline}
  S_{\sigma} = -\frac{1}{2}\int\frac{d^2qd\omega}{(2\pi)^2}\bigg[
  {\bf \sigma}^+({\bf q},\omega)\cdot\big({\bf V}^+\big)^{-1}
  \cdot{\bf \sigma}^+(-{\bf q}.-\omega) \\
 + V^-_2\big|\sigma^-_2\big|^2\bigg]
\end{multline}
where ${\bf \sigma}^+ = \big(\sigma_0^+,\sigma_2^+\big)$ and
\begin{equation}
  \big({\bf V}^+\big)^{-1} =
\begin{pmatrix}
  2\big(\chi_0^0-\frac{1}{F_0}\big) & 2\chi_2^0\\
  2\chi_2^0 & \chi_0^0 + \chi_4^0 -\frac{1}{F_2}
\end{pmatrix}
\end{equation}
As a result, $V^-_2$ is completely unaffected by the presence of
$F_0$ due to the fact that there is no $\sigma_0^-$ field and
opposite ``signs'' decouple.

Again, we may write this expression as a function of $x=K_0(s)$ 
and taking the inverse, we find
\begin{equation}
  {\bf V}^+ = \frac{1}{x^3+ax^2+bx+c}{\bf W}(x)\label{eq:Vmatrix}
\end{equation}
where
\begin{multline}
  {\bf W}(x) = \frac{1}{2\delta_0+\delta_2 - 4}\times\\
\begin{pmatrix}
  -x^3+\big(\delta_2/2\big)x^2 -(1-\delta_2)x + \delta_2/2 & x(1-x^2)\\
  x(1-x^2) & \delta_0(1+x)^2
\end{pmatrix}
\end{multline}
and
\begin{equation}
  a = \frac{(2-\delta_0)\delta_2}{2\delta_0\!+\!\delta_2\!-\!4},
  \quad
  b = \frac{2\delta_0+\delta_2-2\delta_0\delta_2}
  {2\delta_0\!+\!\delta_2\!-\!4},
  \quad
  c = \frac{-\delta_0\delta_2}{2\delta_0\!+\!\delta_2\!-\!4}
\end{equation}
with $\delta_0=1+1/F_0$ and $\delta_2=1+1/F_2$. Again we find that
the polynomial is cubic. Hence, the addition of $F_0$ only
complicates the algebra, but not the general structure of the
solution. We therefore continue as in Appendix \ref{ap:part}.

Solving the cubic equation in \eqref{eq:Vmatrix} leads to an
algebraically complicated result. However, it simplifies near the
nematic QCP and we find to lowest order in $\delta_2$:
\begin{equation}
  \delta_a^+ = i\sqrt{\frac{\delta_0}{\delta_0-2}},\quad
  \delta_b^+ =-i\sqrt{\frac{\delta_0}{\delta_0-2}},\quad
  \delta_c^+ = \delta_2/2
\end{equation}
which return to our previous result if we let $F_0\to0$ or
$\delta_0\to\infty$. Notice that, to lowest order $\delta_c^+$ is
independent of $F_0$, as it should since the s-wave mode is
non-critical and this ``pole'' precisely represents the critical behavior.
Expanding in partial fractions leads to the expression
\begin{equation}
  {\bf V} = \frac{1}{N(0)}\sum_{\beta=a,b,c}
  \frac{{\bf Z}_\beta^+}{\delta_\beta-K_0(s)}
\end{equation}
with residue matrices
\begin{equation}
\begin{split}
{\bf Z}_a^+ &= \frac{i}{2}\sqrt{\frac{\delta_0}{\delta_0-2}}
\begin{pmatrix}
  \frac{1}{\delta_0\big(\delta_0-2\big)} &
    \frac{\delta_0-1}{\delta_0\big(\delta_0-2\big)}\\
  \frac{\delta_0-1}{\delta_0\big(\delta_0-2\big)} &
    1 +\frac{i}{\sqrt{\delta_0\big(\delta_0-2\big)}}
\end{pmatrix}\\
{\bf Z}_b^+ &= \frac{-i}{2}\sqrt{\frac{\delta_0}{\delta_0-2}}
\begin{pmatrix}
  \frac{1}{\delta_0\big(\delta_0-2\big)} &
    \frac{\delta_0-1}{\delta_0\big(\delta_0-2\big)}\\
  \frac{\delta_0-1}{\delta_0\big(\delta_0-2\big)} &
    1 -\frac{i}{\sqrt{\delta_0\big(\delta_0-2\big)}}
\end{pmatrix}\\
{\bf Z}_c^+ &= \frac{1}{2}
\begin{pmatrix}
  0 & 0\\
  0 & 1
\end{pmatrix}
\end{split}
\end{equation}
which again returns to the previous result if we let $F_0\to 0$.
Notice that ${\bf Z}_c^+$ is independent of $F_0$ to this order in
$\delta_2$. Consequently, the critical behavior of the fermions,
\eqref{eq:nematicZ}, is unaffected by the inclusion of a
non-critical mode such as $F_0$.

\section{Equal-Position Boson Propagator}
\label{ap:equalt}

\begin{table}[t]
\newcolumntype{x}{>{\centering\arraybackslash$}X<{$}}
\newcolumntype{P}{>{\centering\arraybackslash$}p{0.2\columnwidth}<{$}}
\begin{tabularx}{\columnwidth}{P x x}
                                                                                  \hline\\
                      & \lim_{\{\nu,u_n\}\to0^-}  & \lim_{\{\nu,u_n\}\to0^+} \\\\\hline\\
I_{\nu}(\nu)          & -0.0053\ln(\nu)/\nu^{1/3} & 0.0053\ln(\nu)/\nu^{1/3} \\\\\hline\\
I'_{\nu}(\nu)         & -0.0018\ln^2(\nu)         & -0.0018\ln^2(\nu)        \\\\\hline\\
{I'_{\nu}}^{(1)}(\nu) & 0.0077/\nu^{1/3}          & -0.0077/\nu^{1/3}        \\\\\hline\\
I_u(u_n)              & 0.0280/u_n^{1/3}          & 0.0140/u_n^{1/3}         \\\\\hline\\
I'_u(u_n)             & 0.0052\ln^2(u_n)          & -0.0052\ln^2(u_n)        \\\\\hline
\end{tabularx}
\caption{Low frequency / long wavelength limit of the required integrals}
\label{tb:ints}
\end{table}

Here we are interested in the quantity
\begin{multline}
  G_{B(S,S)}({\bf 0},t) = G_{B(S,S)}^0({\bf 0},t) +\\
    i\int\frac{d^2kd\varepsilon}{(2\pi)^3}G^0_{F(S)}(k,\varepsilon) 
    V_{S,S}(k,\varepsilon)
    G^0_{F(S)}(k,\varepsilon)
    \left(e^{-i\varepsilon t}-1\right)
\end{multline}
the equal position part of Eq. \eqref{eq:GB}. In particular, we
focus on the second term, labeling it $\ln Z_S({\bf 0},t)$. We
begin by writing the interaction in terms of its spectral
function:
\begin{equation}
  V_{S,S}(k,\varepsilon) = \int \frac{d\varepsilon'}{2\pi}
    \frac{B(k,\varepsilon')}
    {\varepsilon - \varepsilon' +i\epsilon'\sign\varepsilon'}
\end{equation}
where 
\begin{multline}
B({\bf k},\omega)=B_2^+(k,\omega) \cos^2\big(2(\theta_S-\phi)\big) \\ 
									+ B_2^-(k,\omega) \sin^2\big(2(\theta_S-\phi)\big)
\end{multline}
and $\phi$ the direction of ${\bf k}$. Note: in the appropriate scaling within
a patch, $\phi\to\theta_S+\pi/2$ so that the cosine factor scales to $1$ while the sine factor scales to $0$ driving the $B_2^-(k,\omega)$ contribution irrelevant (this has been checked explicitely).

This allows us to do the $\varepsilon$-integration immediately and
rewrite our expression in a more physical form. The result is
(after letting $\epsilon'\to0$):
\begin{multline}\label{eq:lnZsres}
  \ln Z_S({\bf 0},|t|) = \\\int_{-\Lambda/2}^{\Lambda/2} \frac{dk_t}{2\pi}
      \int_{-\lambda/2}^{\lambda/2} \frac{dk_n}{2\pi}
      \int_0^{\infty} 
      \frac{d\varepsilon}{2\pi} B(k,\varepsilon) R_1(k_n,\varepsilon,|t|)
      +\\
      \int_{-\Lambda/2}^{\Lambda/2} \frac{dk_t}{2\pi}
        \int_{0}^{\lambda/2} \frac{dk_n}{2\pi}
        \int_{-\infty}^{\infty} \frac{d\varepsilon}{2\pi} B(k,\varepsilon)
        R_2(k_n,\varepsilon,|t|)
\end{multline}
which we have written directly in terms of the patch coordinates. 
The residue of the single pole is:
\begin{align}
  R_1(k_n,\varepsilon,|t|) &=
    \frac{e^{-i\varepsilon |t|}-1}{\left(\varepsilon-
      v_Fk_n+i\epsilon\sign \varepsilon\right)^2}\\
   &=\frac{1}{v_F}\frac{d}{dk_n}
   \left[\frac{e^{-i\varepsilon |t|}-1}{\varepsilon -
      v_Fk_n+i\epsilon\sign \varepsilon}\right]
\end{align}
and the residue of the double pole is
\begin{equation}
  R_2(k_n,\varepsilon,|t|) = -\frac{1}{v_F}\frac{d}{dk_n}\left[
    \frac{e^{-ik_n|t|}-1}{\varepsilon -v_Fk_n+i\epsilon\sign\varepsilon}\right]
\end{equation}
so that it is clear that an integration-by-parts removes the
double pole all together, leaving us with an integral only over a
single pole. One can view this as a cancelation of the double
pole's contribution to the integral and expect the result to be
less singular than the equal-time case.

Using the symmetries of the spectral function, we may rewrite 
Eq. \eqref{eq:lnZsres} as
\begin{multline}
  \ln Z_S({\bf 0},|t|) = \int_{-\infty}^{\infty} \frac{dk_t}{2\pi}
      \int_{0}^{\infty} \frac{dk_n}{2\pi}
      \int_0^{\infty} \frac{d\varepsilon}{2\pi} 
      B(k,\varepsilon) \times \\
      \bigg[
    R_1(k_n,\varepsilon,|t|) + R_1(-k_n,\varepsilon,|t|) + 
    R_2(k_n,\varepsilon,|t|) -
    R_2(k_n,-\varepsilon,|t|)\bigg]\\
\end{multline}
Performing the integration by parts, and a quick change of
variables allows us to separate the time dependence in terms of
the following integrals:
\begin{widetext}
\begin{multline}
  \ln Z_S({\bf 0},|t|) = \frac{1}{N(0)v_F\sqrt{\kappa}} \Bigg[
    -2\int_0^{\infty} d\nu\bigg(I'_{\nu}(-\nu) -
      \frac{1}{\lambda\sqrt{\kappa}}
      \big(I_{\nu}(-\nu)+{I'_{\nu}}^{(1)}(-\nu)\big)\bigg)+\\
    \int_0^{\infty} du_n\bigg(
      I_{u}'(-u_n) +I_{u}'(u_n) -
        \frac{1}{\lambda\sqrt{\kappa}}\big(I_{u}(-u_n)+I_{u}(u_n)+
        u_nI_{u}'(-u_n)+u_nI_{u}'(u_n)\big)
      \bigg) e^{-i u_n |\bar t|}\\
   +  \int_0^{\infty} d\nu\bigg(
      I_{\nu}'(-\nu) - I_{\nu}'(\nu) -
        \frac{1}{\lambda\sqrt{\kappa}}\big(I_{\nu}(-\nu)-I_{\nu}(\nu)+
        {I'_{\nu}}^{(1)}(-\nu)-{I'_{\nu}}^{(1)}(\nu)\big)
      \bigg) e^{-i \nu |\bar t|}
\bigg]
\end{multline}
\end{widetext}
where we have sent the cutoffs to infinity, keeping the lowest
order correction in $\lambda$ (having checked that higher orders
are non-singular), and where the various integrals are
\begin{align}
  I_{\nu}(\nu)          &= \frac{1}{4\pi^3}{\mathcal P}\int_0^{\infty}du_t
                           \int_0^{\infty} du_n \frac{B(u,\nu)}{\nu - u_n}\\
  I'_{\nu}(\nu)         &= \frac{1}{4\pi^3}{\mathcal P}\int_0^{\infty}du_t
                           \int_0^{\infty} du_n \frac{\frac{d}{du_n}B(u,\nu)}{\nu - u_n}\\
  {I'_{\nu}}^{(1)}(\nu) &= \frac{1}{4\pi^3}{\mathcal P}\int_0^{\infty}du_t
                        \int_0^{\infty} du_n \frac{u_n\frac{d}{du_n}B(u,\nu)}{\nu - u_n}\\
  I_{u}(u_n)   &= \frac{1}{4\pi^3}{\mathcal P}\int_0^{\infty}du_t\int_0^\infty d\nu
                  \frac{B(u,\nu)}{\nu - u_n} \\
  I'_{u}(u_n)  &= \frac{1}{4\pi^3}{\mathcal P}\int_0^{\infty}du_t\int_0^\infty d\nu
                  \frac{\frac{d}{du_n}B(u,\nu)}{\nu - u_n}
\end{align}
In this latest form of $\ln Z_S$, we have defined ${\bf u} = {\bf
k}\sqrt{\kappa}$, $\nu = \varepsilon \sqrt{\kappa}/v_F$, $\bar t =
v_F t/\sqrt{\kappa}$ due to the form of the spectral function at a
nematic instability:
\begin{equation}
  B(k,\varepsilon) = 
  \frac{\varepsilon kv_F}{\varepsilon^2 + \kappa^2 v_F^2 k^6/4} =
    \frac{\nu u}{\nu^2 + u^6/4}
\end{equation}
where we have only included the effects of the overdamped mode since it is responsible for the leading singular behavior. Here we  have replaced the angular factor $\cos^2(2\theta_S)$ by a constant of order unity. (This is consistent since we will be interested in the angular-averaged fermion Green function and the angular factor in the exponent vanishes only on a set of measure zero.) 

Table \ref{tb:ints} shows the low frequency-long wavelength
limit of these integrals computed numerically. Two basic forms
emerge: a $\ln^2(\nu)$ divergence and a $\nu^{-1/3}$ power-law
divergence (with logarithmic corrections) each form occurring at
different scales. However, the $\ln^2(\nu)$ contributions all
vanish. With these limits in mind, we perform the final integral
and obtain
\begin{equation}
  Z_S({\bf 0},|t|) = 
  C_P \exp\bigg\{-A(1+\sqrt{3}i)\frac{\ln (b|\bar t|)}{|\bar t|^{2/3}}\bigg\}
\end{equation}
valid for $|\bar t|\gg 1$. In this expression,
\begin{align}
\label{eq:CP}
  C_P &= \exp\bigg\{ \frac{1}{N(0)v_F\sqrt{\kappa}}
          \left(0.0476+\frac{0.0391}{\lambda\sqrt{\kappa}}+
	  \ldots\right)\bigg\}>1\\
	  \label{eq:A}
  A &= \frac{0.00724}{N(0)v_F\kappa\lambda}
       \propto \frac{1}{k_F\lambda\kappa} \\
       \label{eq:b}
  b &= -4.95+1.51i
\end{align}
where the numbers calculated for $C_P$ were also computed numerically.


\begin{thebibliography}{54}
\bibitem[{com({\natexlab{a}})}]{comment0}
\bibinfo{note}{{A p-wave genaralization of the nematic (d-wave) state has been
  discussed by Wu and Zhang\cite{Congjun04}, by Varma and Zhu \cite{Varma05},
  and by Hirsch\cite{Hirsch90}}}.
\bibitem[{com({\natexlab{b}})}]{comment1}
\bibinfo{note}{{The lattice version of this problem, the spontaneous breaking
  of a point group symmetry, has also been discussed
  recently.\cite{Halboth00,Yamase00,Hankyevich02,Metzner03,
  Kee03,Neumayr03,Kivelson04,Pryadko04,Vadim04,Vadim05}}}.
\end{thebibliography}

\end{document}